\newcommand{\AP}[3]{Ann.\ Phys.\ {\bf #1},\ #2 (#3)}
\newcommand{\NPA}[3]{Nucl.\ Phys.\ {\bf A#1},\ #2 (#3)}
\newcommand{\NPB}[3]{Nucl.\ Phys.\ {\bf B#1},\ #2 (#3)}

\newcommand{\PLB}[3]{Phys.\ Lett.\ B\ {\bf #1},\ #2 (#3)}
\newcommand{\PR}[3]{Phys.\ Rep.\ {\bf #1},\ #2 (#3)}
\newcommand{\PRL}[3]{Phys.\ Rev.\ Lett.\ {\bf #1},\ #2 (#3)}

\newcommand{\PRC}[3]{Phys.\ Rev.\ C\ {\bf #1},\ #2 (#3)}
\newcommand{\PRD}[3]{Phys.\ Rev.\ D\ {\bf #1},\ #2 (#3)}
\newcommand{\JPG}[3]{J.\ Phys.\ G\ {\bf #1},\ #2 (#3)}

\newcommand{\ZPC}[3]{Z.\ Phys.\ C\ {\bf #1},\ #2 (#3)}

\newcommand{\EPJA}[3]{Eur.\ Phys.\ J.\ A\ {\bf #1},\ #2 (#3)}
\newcommand{\PTP}[3]{Prog.\ Theo.\ Phys.\ {\bf #1},\ #2 (#3)}

\newcommand{\diracslash}[1]{#1\llap{/\kern2pt}}

\newcommand{\be}{\begin{equation}}
\newcommand{\ee}{\end{equation}}
\newcommand{\bea}{\begin{eqnarray}}
\newcommand{\eea}{\end{eqnarray}}
\newcommand{\ba}[1]{\begin{array}{#1}}
\newcommand{\ea}{\end{array}}

\documentclass[prd,aps,floats,nofootinbib,tightenlines,showpacs]{revtex4}
\usepackage{graphicx}
\begin{document}

\title{Color superconducting 2SC+s quark matter \\

and gapless modes at finite temperatures}
\author{Amruta Mishra}
\email{mishra@th.physik.uni-frankfurt.de}
\affiliation{Department of Physics, Indian Institute of Technology, New 
Delhi-110016,India}
\email{amruta@physics.iitd.ac.in}
\affiliation{Institute f\"ur Theoretische Physik, 
Universit\"at Frankfurt, D-60054 Frankfurt, Germany}

\author{Hiranmaya Mishra}
\email{hm@prl.ernet.in}

\affiliation{Theory Division, Physical Research Laboratory,
Navrangpura, Ahmedabad 380 009, India}

\date{\today} 

\def\be{\begin{equation}}
\def\ee{\end{equation}}
\def\bearr{\begin{eqnarray}}
\def\eearr{\end{eqnarray}}
\def\zbf#1{{\bf {#1}}}
\def\bfm#1{\mbox{\boldmath $#1$}}
\def\hf{\frac{1}{2}}
\begin{abstract}
We investigate the phase diagram of color superconducting quark matter 
with strange quarks (2SC+s quark matter) in beta equliibrium at zero as
well as finite temperatures within a Nambu-Jona-Lasinio model. The 
variational method as used here allows us to investigate simultaneous
formation of condensates in quark--antiquark as well as in diquark channels.
Color and electric charge neutrality conditions are imposed in the 
calculation of the thermodynamic potential. Medium dependance of strange 
quark mass plays a sensitve role in maintaining charge neutrality conditions.
At zero temperature the system goes from gapless phase to usual BCS phase
through an intermediate normal phase as density is increased. The gapless
modes show a smooth behaviour with respect to temperature vanishing above a 
critical temperature which is larger than the BCS transition temperature. We 
observe a sharp transition from gapless superconducting phase to the BCS 
phase as density is increased for the color neutral matter at zero temperature. 
As temperature is increased this however becomes a smooth transition.
\end{abstract}

\pacs{12.38.Mh, 24.85.+p} 

\maketitle

 \section{Introduction}

Color superconductivity has become a compelling topic in QCD during 
the last few years. At sufficiently high baryon densities, when nucleons 
get converted to quark matter, the resulting quark matter is in one kind 
or the other of the many different possible color
superconducting phases at low enough temperatures \cite{review}. The rich
phase structure is essentially due to the fact that the quark quark interaction
is not only strong and attractive in many channels but also many degrees of
freedom are possible for quarks like color, flavor and spin so that
various kinds of BCS pairing are possible.  Studying properties of color
superconducting phases in heavy ion collision experiments seems unlikely
in the present accelerators as one cannot avoid producing large entropy
per baryon in heavy ion collisions and hence cannot produce the dense and
cold environment that is needed to supoort the formation of superconducting 
phases.  However, in the future acclerator facility planned at GSI for 
compressed baryonic matter experiments, one possibly can hope for 
observing fluctuations signifying precursory phenomena of color 
superconducting phase \cite{kunihiroo}.

        On the other hand, it is natural to expect some color suprconducting
phase to exist in the core of compact stars where the densities are above
nuclear matter densities and temperatures are of the order of tens of keV.
However, to consider quark matter for neutron stars, color and
electrical charge neutrality conditions need to be imposed for the bulk
quark matter. Such an attempt has been made 
in Ref.\cite{krisch} as well as in Ref. \cite{reddy} where the 
lighter up and down quarks form the two flavor color superconducting (2SC)
matter while the strange quarks do not participate in pairing. 
A model independent analysis was done  in Ref. \cite{krisch}
that is valid in the limit $m_s<<\mu$ and $\Delta\sim m_s^2/\mu$, where,
 $\Delta$ is the pairing gap and $\mu$ is the quark chemical potential.
It has been shown, based upon comparison of
free energies that such a two flavor color superconducting phase
would be absent in the core of neutron stars \cite{krisch}.
Within Nambu Jona-Lasinio (NJL) model in Ref. \cite{reddy}
it has been argued that such conclusions are consistent except for a small
window in  density range where superconducting phase is 
possible. There have also been studies to include the possibility of mixed 
phases \cite{bubmix} of superconducting matter demanding neutral
matter on the average. Recently it was observed that imposition of 
neutrality conditions lead to pairing of quarks with different fermi momenta
giving rise to gapless modes \cite{igor,hmam}. Within a Nambu JonaLasinio model
the two flavor superconducting quark matter (2SC)  was shown to 
exhibit gapless modes (g2SC) arising due to difference in the fermi momenta 
of the pairing quarks when charge and color neutrality conditions are
imposed. 
Superconducting quark matter with strange quarks (2SC+s) was shown to 
exhibit gapless superconductivity (g2SC+s) within a window of about
80 MeV in baryon chemical potential \cite{hmam}. The number densities 
of the strange quarks were seen to be as large as about 40$\%$ of the 
number density of e.g. the  up quarks in the gapless phase \cite{hmam}.
Temperature effects on the gapless modes were also studied for two flavor
quark matter in Ref.\cite{igorr}.

We had applied a different approach to study the problem in 
Ref. \cite{hmam,npa}. We considered a variational approach with 
an explicit assumption for the ground state having both quark--antiquark 
as well as diquark condensates. The
actual calculations are carried out for the NJL model with a vector
interaction such that the minimisation of the free energy density
determines which condensate will exist at what density.
Charge neutrality conditions were introduced through the introduction
of  appropriate chemical potentials.  We note here that
 possibility of diquark condensates along with quark--antiquark 
condensates has also been considered in Ref.
\cite{{berges},{mei},{blaschke},{kunihiro}}.

In the present work we generalise the variational approach of Ref. \cite{hmam} 
to study the color superconductivity as well as the chiral symmetry
breaking  to include the effects of temperatures. This will be
particularly relevant for the physics of  proto neutron stars.
The gapless modes in the two flavor color superconductivity were 
investigated at finite temperatures \cite{igorr}.
However, in the present work we take the analysis further in the sense 
that the diquark and the quark--antiquark condensates are uniquely 
calculated selfconsistently and elaborate the transition from gapless phase 
and BCS phase as the density is increased for different temperatures. 
Further, since the realisation of color and charge neutrality is nontrivial
for physically relevant values of strange quark mass, we retain
all orders in $m_s$ in the calculation of the thermodynamic potential.
This of course is important to understand the phase structure of quark matter
at densities relevant to the interior of neutron stars. The temperature 
dependence and the nature of the gapless modes 
as a function of the baryon density shall be 
particularly relevant for cooling of the neutron stars through 
neutrino emission \cite{prasanth}.

We organize the paper as follows. In the next subsection we discuss the ansatz
state with the quark--antiquark as well as the diquark condensates.
In section \ref{evaluation}  we consider the Nambu Jona-Lasinio model
 Hamiltonian and calculate its expectation value with respect to the
 ansatz state to compute
the thermodynamic potential. We minimise the thermodynamic potential
to calculate all the ansatz functions and the resulting mass as well as
superconducting gap equations here. In section \ref{results} we discuss
the results of the present investigation. Finally we summarise 
and conclude in section \ref{summary}.
\subsection{ An ansatz for the ground state}

To make the notations clear, let us note first the quark field operator
expansion in momentum space given as
 \cite{hmnj,amspm}
\begin{eqnarray}
\psi (\zbf x )\equiv &&\frac{1}{(2\pi)^{3/2}}\int \tilde\psi(\zbf k)
e^{i\zbf k\cdot\zbf x}d\zbf k \nonumber\\ 
=&&\frac{1}{(2\pi)^{3/2}}\int \left[U_0(\zbf k)q^0(\zbf k )
+V_0(-\zbf k)\tilde q^0(-\zbf k )\right]e^{i\zbf k\cdot \zbf x}d \zbf k,
\label{psiexp}
\end{eqnarray}
where,
\begin{eqnarray}
U_0(\zbf k )=&&\left(\begin{array}{c}\cos(\frac{\phi^0}{2})\\
\zbf \sigma \cdot \hat k \sin(\frac{\phi^0}{2})
\end{array}\right),\;\;
V_0(-\zbf k )=
\left( \begin{array}{c} -\zbf \sigma \cdot \hat k  \sin(\frac{\phi^0}{2})
\\ \cos(\frac{\phi^0}{2})\end{array}
\right).
\label{uv0}
\end{eqnarray}
The operators $q^0$ and $\tilde q^0$ are the
two component particle annihilation and antiparticle creation operators 
respectively which annihilate or create quanta acting upon
the perturbative or the chiral vacuum $|0\rangle$. We have suppressed 
here the color and flavor indices of the quark field operators. The function 
$\phi^0(\zbf k)$ in the spinors in Eq.(\ref{uv0}) are given as 
$\cot{\phi_i^0}=m_i/|\zbf k|$,  for free massive fermion fields, 
$i$ being the flavor index. For massless fields
$\phi^0(|\zbf k|)=\pi/2$.

We now write down the ansatz for the variational state as a squeezed
coherent state involving quark antiquark as well as diquark condensates
as given by
\cite{hmam,npa} 
\begin{equation} 
|\Omega\rangle= {\cal U}_d|vac\rangle={\cal U}_d{\cal U}_Q|0\rangle,
\label{u0}
\end{equation} 

Here, ${\cal U}_Q$ and ${\cal U}_d$ are unitary operators creating
quark--antiquark and diquark pairs respectively. Explicitly,
\begin{equation}
{\cal U}_Q=\exp\left (
\int q^{0i}(\zbf k)^\dagger(\bfm {\sigma }\cdot\zbf k) h_i(\zbf k)
 \tilde q^{0i} (\zbf k)d\zbf k-h.c.\right ).
\label{u1}
\end{equation}
 
\noindent In the above, $h_i(\zbf k)$  is a
real function of $|\zbf k|$ which describes vacuum realignment for
chiral symmetry breaking for quarks of a given flavor $i$. We shall take 
the condensate function $h(\zbf k)$ to be the same for u and d quarks and
$h_3(\zbf k)$ as the chiral condensate function for the s-quark.
Clearly, a nontrivial $h_i(\zbf k)$ shall break chiral
symmetry. Summation over three colors and three flavors is understood in the
exponent of ${\cal U}_Q$
in Eq. (\ref{u1}). Similarly, the unitary operator ${\cal U}_d$
describing diquark condensates is given as

\begin{equation}
{\cal U}_d=\exp(B_d^\dagger-B_d)
\label{omg}
\end{equation}
where, $B_d^\dagger$ is the pair creation operator as given by
\begin{equation}
{B}_d ^\dagger=\int \left[q_r^{ia}(\zbf k)^\dagger
r f^{ia}(\zbf k) q_{-r}^{jb}(-\zbf k)^\dagger
\epsilon_{ij}\epsilon_{3ab}
+\tilde q_r^{ia}(\zbf k)
r f_1^{ia}(\zbf k) \tilde q_{-r}^{jb}(-\zbf k)
\epsilon_{ij}\epsilon_{3ab}\right]
d\zbf k.
\label{bd}
\end{equation}
\noindent 
In the above, $i,j$ are flavor indices, $a,b$ are the
color indices and $r(=\pm 1/2) $ is the spin index.
The operators $q(\zbf k)$ are related to $q^0(\zbf k)$ through
the transformation $q(\zbf k)={\cal U}_Q q^0(\zbf k){\cal U}_Q^{-1}$.
As noted earlier we 
shall have the quarks of colors red and green ($a$=1,2) and 
flavors u,d ($i$=1,2) taking part in diquark condensation .
The blue quarks (a=3) do not take part in diquark condensation.
We have also introduced  here (color, flavor dependent) functions
 $f^{ia}(\zbf{k})$ and
$f^{ia}_1(\zbf k)$ respectively for the diquark and diantiquark
channels. As may be noted the state constructed in Eq.(\ref{omg}) is spin
singlet and is antisymmetric in both color and flavor.
Clearly, by construction $f^{ia}(\zbf k)=f^{jb}(\zbf k)$ with $i\neq j$ and
$a\neq b$. The ansatz for the ground state has been written down 
keeping quark--antiquark condensates for the three flavors and diquark 
condensates for two flavors and two colors.

Finally, to include the effects of temperature and density we next write
 down the state at finite temperature and density 
$|\Omega(\beta,\mu)\rangle$  taking
a thermal Bogoliubov transformation over the state $|\Omega\rangle$ 
using thermofield dynamics (TFD) as described in ref.s \cite{tfd,amph4}.
We then have,
\begin{equation} 
|\Omega(\beta,\mu)\rangle={\cal U}_{\beta,\mu}|\Omega\rangle={\cal U}_{\beta,\mu}
{\cal U}_d{\cal U}_Q |0\rangle.
\label{ubt}
\end{equation} 
where ${\cal U}_{\beta,\mu}$ is given as
\begin{equation}
{\cal U}_{\beta,\mu}=e^{{\cal B}^{\dagger}(\beta,\mu)-{\cal B}(\beta,\mu)},
\label{ubm}
\end{equation}
with 
\begin{equation}
{\cal B}^\dagger(\beta,\mu)=\int \Big [
q_I^\prime (\zbf k)^\dagger \theta_-(\zbf k, \beta,\mu)
\underline q_I^{\prime} (\zbf k)^\dagger +
\tilde q_I^\prime (\zbf k) \theta_+(\zbf k, \beta,\mu)
\underline { \tilde q}_I^{\prime} (\zbf k)\Big ] d\zbf k.
\label{bth}
\end{equation}
In Eq.(\ref{bth}) the ansatz functions $\theta_{\pm}(\zbf k,\beta,\mu)$
will be related to quark and antiquark distributions and the underlined
operators are the operators in the extended Hilbert space associated with
thermal doubling in TFD method. In Eq.(\ref{bth}) we have suppressed
the color and flavor indices on the quarks as well as the functions
$\theta(\zbf k,\beta,\mu)$.
 All the  functions in the ansatz in Eq.(\ref{ubt})
are to be obtained by minimising the
thermodynamic potential. This will involve an 
assumption about the effective 
Hamiltonian. We shall carry out this minimisation
in the next section.

\section{Evaluation of thermodynamic potential and gap equations }
\label{evaluation}

Since we shall be dealing with non--asymptotic densities we can not use the
weak coupling method. For this reason we shall be working 
in a model in which the interaction among the quarks is simplified
-- while still respecting the symmetries of QCD. One natural choice 
is to model the interaction between the quarks using a four fermion point 
interaction -- namely, the Nambu Jona Lasinio model.

The Hamiltonian is given as
\be
{\cal H}=\sum_{i,a}\psi^{ia \dagger}(-i\bfm \alpha \cdot \bfm \nabla
+\gamma^0 m_i )\psi^{ia}
-G_s\sum_{A=0}^8\left[(\bar\psi\lambda^A\psi)^2-
(\bar\psi\gamma^5\lambda^A\psi)^2\right]
-G_D(\bar\psi\gamma^5\epsilon\epsilon^b\psi^C)(\bar\psi^C\gamma^5\epsilon\epsilon^b\psi)
\label{ham}
\ee
where $\psi ^{i,a}$ denotes a quark field of color  $a$ and flavor $i$,
and $m_i$ is the current quark mass. As noted earlier, we  shall assume isospin
symmetry with $m_u$=$m_d$.  The interaction term in Eq.(\ref{ham})
is written down assuming that the underlying gluonic degrees of freedom 
can be frozen into point like effective interactions between the quarks.
In Eq.(\ref{ham}) $\lambda^A$, $A=1,\cdots 8$ denote the Gellman matrices
acting in the flavor space and
$\lambda^0 = \sqrt{\frac{2}{3}}\,1\hspace{-1.5mm}1_f$,
$1\hspace{-1.5mm}1_f$ as the unit matrix in the flavor space.
The point interaction produces short distance singularities and 
to regulate the integrals we shall restrict the phase space inside 
the sphere $|\zbf p|< \Lambda$ -- the ultraviolet cutoff in the NJL model.
This form of Lagrangian can arise e.g. by Fiertz
transformation of a four point current current interaction having 
quantum numbers of single-gluon exchange \cite{ebert}. In that case 
the diquark coupling $G_D$ is related to the scalar coupling
as $G_D=0.75G_s$.
In the study of meson spectrum in this model \cite{rehberg,lutz} one
adds a six point t'Hooft interaction which breaks $U(1)_A$ symmetry.
Although it is straightforward to include such a term, we neglect it here
for the sake of simplicity. This can have some qualitative consequences
as we shall discuss later. 

Using the fact that the variational ansatz state in Eq.(\ref{ubt}) 
arises from successive Bogoliubov
transformations one can calculate the expectation values
of various operators \cite{hmam}.
These expressions are used to calculate thermal 
expectation value of the Hamiltonian to compute the thermodynamic potential. 
A straightforward, but cumbersome calculation yields 
\be
\epsilon = \langle H \rangle = T +V_S +V_D 
\ee
arising from the kinetic energy part and the scalar and diquark interaction
terms of the Hamiltonian respectively. Explicitly,
the kinetic energy part in Eq.(\ref{ham}) is given as 
\bearr
T & \equiv &
 \langle \Omega(\beta,\mu)|
\psi^\dagger
(-i\bfm \alpha \cdot \bfm \nabla +\gamma^0 m_i )\psi
| \Omega(\beta,\mu)\rangle \nonumber \\
 & = & -\frac{2}{(2\pi)^3}\sum_{i=1}^3\sum_{a=1}^3
\int d \zbf k (|\zbf k|\sin \phi^i+m_i\cos\phi_i)\left(1-\cos 2h_i(\zbf k)(1-F^{ia}-F_1^{ia})\right),
\label{tren}
\eearr
where, $F^{ia}$ and $F_1^{ia}$ are defined as
\begin{equation}
F^{ia}(\zbf k)=\sin^2\theta^{ia}_-(\zbf k)+\sin^2 f^{ia}\left(C_-^{ia}(\zbf k)-
\sin^2\theta_-^{ia}(\zbf k)\right)\left(1-\delta^{a3}\right),
\label{fkb}
\end{equation}
and,
\begin{equation}
F_1^{ia}(\zbf k)=\sin^2\theta^{ia}_+(\zbf k)+\sin^2 f_1^{ia}
\left(\cos^2\theta_+^{ia}(\zbf k)- S_+^{ia}(\zbf k)\right)
\left(1-\delta^{a3}\right).
\label{f1kb}
\end{equation}
Here, we have defined $C_-^{ia}=
|\epsilon^{ii'}\epsilon^{aa'}|\cos^2\theta_-^{i'a'}$ and
$S_+^{ia}=|\epsilon^{ii'}\epsilon^{aa'}|\sin^2\theta_+^{i'a'}$.
The $\delta^{a3}$ term indicates that the third color does not take part in 
diquark condensation. 

The contribution from the scalar interaction  term in Eq.(\ref{ham})
turns out to be
\begin{equation}
{V_S}\equiv -G_s \langle \Omega(\beta,\mu)|
\sum_{A=0}^8\left[(\bar\psi\lambda^A\psi)^2-
(\bar\psi\gamma^5\lambda^A\psi)^2\right]
| \Omega(\beta,\mu)\rangle
=-8 G_S\sum_{i=1,3}{I_S^i}^2
\label{vs}
\end{equation}

where, 
\be
I_s^i=\sum_{a=1,3}\int d\zbf k \cos\phi^i(1-F^{ia}-F_1^{ia})
\label{is}
\ee
is proportional to $\langle\bar\psi^i\psi^i\rangle$ ( $i$, not summed)
and the function $\phi_i$ is defined in terms of the
quark--antiquark condensate functions as 
 $\phi_i(\zbf k) = \phi_i^0(\zbf k) - 2h_i (\zbf k)$.

In a similar manner, the contribution from the diquark interaction
from Eq.(\ref{ham}) to the energy density is given as
\be
V_D=
-G_D \langle \Omega(\beta,\mu)|
(\bar\psi\gamma^5\epsilon^b\psi^C)(\bar\psi^C\gamma^5\epsilon\epsilon^b\psi)
| \Omega(\beta,\mu)\rangle
=-G_D\left( \sum_{i,j=1,2; a,b=1,2}I^{ia,jb}|\epsilon^{ij}||\epsilon^{3ab}|\right)^2
\label{vd}
\ee
with
\be
I^{ia,jb}=\frac{1}{(2\pi)^3}\int d \zbf k S^{ia,jb}(\zbf k)
\cos\left(\frac{\phi_i-\phi_j}{2}\right)
\label{i3}
\ee

\noindent where, $S^{ia,jb}$ has been defined as
\be
S^{ij,ab}(\zbf k)=\sin\!2f^{ia}(\zbf k)\cos 2\theta^{ia,jb}_-(\zbf k,\beta,\mu)
+\sin\!2f_1^{ia}(\zbf k)\cos 2\theta^{ia,jb}_+(\zbf k,\beta,\mu).
\label{sk}
\ee
\noindent In the above we have defined $\cos 2\theta^{ia,jb}_\pm
=1-\sin^2\theta_\pm^{ia}- \sin^2\theta_\pm^{jb}$, with ${i,j=1,2}$ 
being the flavor indices and $a,b=1,2$ being the color indices and
 $i\neq j$, $a\neq b$.

To calculate the thermodynamic potential we shall have to specify the
chemical potentials relevant for the system.
Here we shall be interested in the form of quark matter that might be present
in compact stars older than few minutes so that 
chemical equilibriation under weak
interaction is there. The relevant chemical potentials in this case then
are the baryon chemical potential $\mu_B=3\mu$, the chemical potential 
$\mu_E$ associated with electromagnetic charge $Q=diag(2/3,-1/3,-1/3)$
in flavor space, 
and the two color electrostatic chemical potentials $\mu_3$ and $\mu_8$ 
corresponding to $U(1)_3\times U(1)_8$ subgroup of the color gauge symmetry
generated by cartan subalgebra $Q_3=diag(1/2,-1/2,0)$ and 
$Q_8=diag(1/3,1/3,-2/3)$ in the color space. Thus the chemical potential
is a diagonal matrix in color and flavor space, and is given by
\be
\mu_{ij,ab}=(\mu\delta_{ij}+Q_{ij}\mu_E)\delta_{ab}
+(Q_{3ab}+Q_{8ab}\mu_8)\delta_{ij}.
\label{muij}
\ee
Here, $i,j$ are flavor indices and $a,b$ are color indices.

The total thermodynamic potential, including the contribution from the
electrons, is then given by
\be
\Omega=T+V_S+V_D-\langle \mu N\rangle-\frac{1}{\beta}s+\Omega_e
\label{Omega}
\ee
where, we have denoted
\be
\langle \mu N\rangle=\langle \psi^{ia\dagger}\mu_{ij,ab}\psi^{jb}\rangle
=2\sum_{i,a}\mu^{ia}I_v^{ia}
\ee
with $\mu^{ia} $ being the chemical potential for the quark of flavor $i$
and color $a$, which can be expressed in terms of the chemical potentials
$\mu$, $\mu_E$, $\mu_3$ and $\mu_8$ using Eq.(\ref{muij}). 
Further
\be
I_v^{ia}=\frac{1}{(2\pi)^3}\int d\zbf k(F^{ia}-F_1^{ia})
\label{iv}
\ee
 is proportional to the number density of quarks of given color
and flavor. The thermodynamic potential for electrons is given as
\be
\Omega_e=
-\frac{\mu_E^4}{12\pi^2}\left(1+2\pi^2 \frac{T^2}{\mu_E^2}\right)
\label{omge}
\ee
where we have taken the electron mass to be zero which suffices for
the system we are considering.

Finally, for the entropy density for the quarks we have \cite{tfd}
\bearr
s & = & -\frac{2}{(2\pi)^3}\sum_{i,a}\int d \zbf k
\Big ( \sin^2\theta^{ia}_-\ln \sin^2\theta^{ia}_-
+\cos^2\theta^{ia}_-\ln \cos^2\theta^{ia}_- \nonumber \\ 
& + & \sin^2\theta^{ia}_+\ln \sin^2\theta^{ia}_+
+\cos^2\theta^{ia}_+\ln \cos^2\theta^{ia}_+\Big ).
\label{ent}
\eearr

Now functional  minimisation  the thermodynamic potential $\Omega$ with respect
 to  the chiral condensate function $h _i (\zbf k)$ leads to
\be
\cot \phi_i(\zbf k)
= \frac{ m_i+8G_sI_s^i}{|\zbf k|}\equiv \frac{M_i}{|\zbf k|}
\label{tan2h}
\ee
\noindent where, $M_i= m_i +8 G_s I^{i}_s$. 
Substituting this back in Eq.(\ref{is}) yields the mass gap equation as
\be
M_j=m_j+ \frac{ 8G_s}{(2\pi)^3}
\int \frac{M_j}{\epsilon_j^2} \sum_{a=1,3}(1-F^{ja}-F^{ja}_1)
 d \zbf k,
\label{mgap}
\ee
where, $\epsilon_j=\sqrt{\zbf k^2+M_j^2}$ being the energy of the
constituent quarks of j-th flavor.
Clearly, the above includes the effect of diquark condensates 
as well as temperature and density through the functions $F$ and $F_1$ 
given in Eq.s (\ref{fkb})
and (\ref{f1kb}) respectively.

As noted earlier, there are only two independent diquark condensate
functions $f^{11}$ and $f^{12}$ corrsponding to the indices u-red and 
u-green respectively. This is due to the fact that the function with 
indices d,green is the same as that with indices u,red and the 
condensate function with indices d-red is the same as that with 
indices u-green.

Minimisation of the thermodynamic potential  $\Omega$ with respect to 
the diquark condensate functions $f^{11}(\zbf k)$ and $f_{12}(\zbf k)$ 
yields 

\be
\tan 2f^{11}(\zbf k)=\frac{\Delta}{\bar \epsilon-\bar \mu_{11}}
\cos(\frac{\phi_1-\phi_2}{2})
\label{tan2f11}
\ee
and
\be
\tan 2f^{12}(\zbf k)=\frac{\Delta}{\bar \epsilon-\bar \mu_{12}}
\cos(\frac{\phi_1-\phi_2}{2})
\label{tan2f12}
\ee
where, we have defined $\Delta= 4 G_D(I^{11,22}+I^{12,21})$ with
$I^{ia,jb}$ as defined in Eq.(\ref{i3}). Further,
in the above $\bar\epsilon=(\epsilon_1+\epsilon_2)/2$, 
$\bar\mu_{11}= (\mu_{11}+\mu_{22})/2$,
$\bar\mu_{12}= (\mu_{12}+\mu_{21})/2$.
It is thus seen that the diquark condensate functions depend upon
the {\em average} energy and the {\em average} chemical potential
of the quarks that condense. We also note here that the 
diquark condensate functions depends upon the masses of the two quarks which
condense through the function $\cos \big ((\phi_1-\phi_2)/2\big )$.
The function $\cos\phi_i=M_i/\epsilon_i$, 
can be different for u,d quarks,
when charge neutrality condition is imposed. 
Such a normalisation factor is always there when
the condensing fermions have different masses as has been noted in Ref.
\cite{aichlin} in the context of CFL phase. 

In an identical manner the di-antiquark condensate functions are 
calculated to be
\be
\tan 2f^{11}_1(\zbf k)=\frac{\Delta}{\bar \epsilon+\bar \mu_{11}}
\cos(\frac{\phi_1-\phi_2}{2})
\label{tan2f1}
\ee
\be
\tan 2f^{12}_1(\zbf k)=\frac{\Delta}{\bar \epsilon+\bar \mu_{12}}
\cos(\frac{\phi_1-\phi_2}{2}).
\label{tan2f112}
\ee

\noindent Substituting  the solutions for the condensate functions 
in the expressions for $I^{ia,jb}$ in Eq.(\ref{i3}) we have the gap equation for $\Delta$ given as
\bearr
\Delta&=&
 4 G_D(I^{11,22}+I^{12,21})\nonumber\\
&=&\frac{4G_D}{(2\pi)^3}\int  d\zbf k\; 
\Delta\left(\frac{\cos 2\theta_-^{12,21}}{\bar\omega_-^{12}}
+\frac{\cos 2\theta_-^{11,22}}{\bar\omega_-^{11}}+
\frac{\cos 2\theta_+^{12,21}}{\bar\omega_+^{12}}+\frac{\cos 2\theta_+^{11,22}
}{\bar\omega_+^{11}}\right)
\cos^2\Big(\frac{\phi_1-\phi_2}{2}\Big)
\label{del}
\eearr

In the above, 
 $\bar\omega_\pm^{ia}
=\sqrt{\Delta^2\cos^2((\phi_1-\phi_2)/2) +\bar\xi_{\pm}^2}$,
${\bar \xi }_{\pm ia}=\bar \epsilon \pm {\bar \mu}^{ia}$ ,and, $\cos 2\theta_\pm^{ia,jb}$
has been defined after Eq.(\ref{sk}). 

Finally, the minimisation of the thermodynamic potential with respect to the
thermal functions $\theta_{\pm}(\zbf k)$ gives
\be
\sin^2\theta_\pm^{ia}=\frac{1}{\exp(\beta\omega_\pm^{ia})+1}
\label{them}
\ee

 \noindent Various $\omega^{ia}$'s are given explicitly as follows.
$$\omega_\pm^{11} =\bar\omega_\pm^{11} +\delta_\epsilon\pm \delta_\mu^{11}$$
$$\omega_\pm^{12} =\bar\omega_\pm^{12} +\delta_\epsilon\pm\delta_\mu^{12}$$
$$\omega_\pm^{21} =\bar\omega_\pm^{12}-\delta_\epsilon\mp\delta_\mu^{12}$$
\be
\omega_\pm^{22} =\bar\omega_\pm^{11}-\delta_\epsilon\mp\delta_\mu^{11}
\label{disps}
\ee

\noindent and, finally,
for the noncondensing colors $\omega_{\pm}^{i3}=\epsilon^i{\pm}\mu^{i3}$.
Also for the strange quark ($i$=3) $\omega_{\pm}^{3a}=\epsilon^3{\pm}\mu^{3a}$,
with $\epsilon_3=\sqrt{\zbf k^2+M_3^2}$.
As already mentioned, the first index refers to flavor and the second 
index refers to color. Here
$\delta _ \epsilon=(\epsilon_1-\epsilon_2)/2$ is half the energy difference
of the two quarks which condense and e.g. $\delta_\mu^{11}=
(\mu_{11}-\mu_{22})/2$, is half the difference of the chemical potentials
of the two quarks which condense.
Further $\bar\omega_\pm^{ia}$ are defined earlier after Eq.(\ref{del}).
 Note that in the absence of imposing the
charge neutrality condition all the four quasi particles will have the
same energy $\bar\omega_-$. It is clear from the
dispersion relations given in Eq.(\ref{disps}) that  
it is possible to have zero modes, i.e., $\omega^{ia}=0$
depending upon the values of $\delta_\epsilon$
and $\delta_\mu$. So, although we shall have nonzero order
parameter $\Delta$, there will be fermionic zero modes or the 
gapless superconducting phase \cite {abrikosov, krischprl}. 
We shall discuss more about it section \ref{results}.

Next, let us focus our attention for the specific case of 
superconducting phase and the chemical potential associated with it. First
let us note that the diquark condensate functions depend upon the 
average of the chemical potentials of the quarks that condense. Since
this is independent of $\mu_3$ we can choose $\mu_3$ to be zero.
In that case, $\mu^{11}=\mu +2/3 \mu_E+\mu_8/\sqrt{3}=\mu^{12}$
and also $\mu^{21}=\mu -1/3 \mu_E+\mu_8/\sqrt{3}=\mu^{22}$.
Thus the chemical potentials of the light flavors become degenerate
for both the colors that take part in condensation. 
This also means that the average chemical potential of both
the condensing quarks are the same  i.e. 
$\bar\mu^{11}=\bar\mu^{12}=\bar\mu=\mu+1/6\mu_E+\mu_8/\sqrt{3}$.
For the same reason we also have ${\delta_\mu}^{12}=\mu_E/2
={\delta_\mu}^{11}
\equiv\delta_\mu$. 

With this condition, it is also clear from Eq (\ref{disps})
that the quasi particle energies  for each flavor becomes degenerate 
for both the colors which take part in condensation. Thus the quasi
particle energies now become 
$\omega_{-1}=\bar\omega_-+\delta_\epsilon -\delta_\mu$
and $\omega_{-2}=\bar\omega_--\delta_\epsilon +\delta_\mu$, 
for u and d quarks respectively. Similarly for the
antiparticles the energies are given as  
$\omega_{+1}=\bar\omega_++\delta_\epsilon +\delta_\mu$
and $\omega_{+2}=\bar\omega_++\delta_\epsilon -\delta_\mu$, 
for u and d quarks.
The gap equation given by Eq.(\ref{del}) then reduces to
\be
\Delta
=\frac{8G_D}{(2\pi)^3}\int d\zbf k 
\Delta
\left[\frac{1}{\bar\omega_-}\left(\cos^2 \theta_-^{1}-\sin^2\theta_2\right)
+\frac{1}{\bar\omega_+}\left( \cos^2\theta_+^{1}-\sin^2\theta_+^{2}\right)
\right]
\cos^2(\frac{\phi_1-\phi_2}{2}),
\label{scgap}
\ee
where,
 $\bar\omega_\pm=\sqrt{\Delta^2\cos^2((\phi_1-\phi_2)/2) +(\bar\epsilon\pm
\bar\mu)^2}$.

Now using these dispersion relations, the mass
gap equation Eq.(\ref{mgap}) and the superconducting gap
equation Eq.(\ref{scgap}), the thermodynamic potential 
given in Eq.(\ref{Omega}) becomes
\be
\Omega=\Omega_{ud}+\Omega_{s}+\Omega_e,
\label{omgt}
\ee
where, $\Omega_{ud}$ is the contribution from the light quarks
and is given as
\bearr
\Omega_{ud}&=& \frac{8}{(2\pi)^3}
\int d\zbf k\left[\sqrt{\zbf k^2+m^2}-
\frac{1}{2}(\bar\omega_-+\bar\omega_+)\right]\nonumber\\
&-& \frac{4}{\beta(2\pi)^3}
\sum_{i=1,2}\int d\zbf k \left[\log(1+\exp (-\beta\omega_{-i})
+\log(1+\exp({-\beta\omega_{+i}})\right]\nonumber\\
&+& \frac{\Delta^2}{4G_D}+\sum_{i=1,2}\frac{(M^i-m)^2}{8G_s}\nonumber\\
&+&\frac{4}{(2\pi)^3}
\int d\zbf k\left[\sqrt{\zbf k^2+m^2}-
\frac{1}{2}(\epsilon_1+\epsilon_2)\right]\nonumber\\
&-& \frac{2}{\beta(2\pi)^3}
\sum_{i=1,2}\int d\zbf k \left[\log(1+\exp({-\beta(\epsilon_i-\mu^{i3})})
+\log(1+\exp({-\beta(\epsilon_i+\mu^{i3})})\right].
\label{omgud}
\eearr
Similarly the contribution from the strange quarks to the 
thermodynamic potential, $\Omega_s$ is given as
\bearr
\Omega_s &=&
\frac{6}{(2\pi)^3}
\int d\zbf k\left[\sqrt{\zbf k^2+m_s^2}-
\sqrt{\zbf k^2+M_s^2}\right]\nonumber\\
&-& \frac{2}{\beta(2\pi)^3}
\sum_{a=1,3}\int d\zbf k \left[\log(1+\exp({-\beta(\epsilon_3-\mu^{3a})})
+\log(1+\exp({-\beta(\epsilon_3+\mu^{3a})})\right]
+\frac{(M_s-m_s)^2}{8G_s}.\nonumber\\
\label{omgs}
\eearr
Finally, the  contribution of the electron to the total 
thermodynamic potentail $\Omega_e$ is as given in Eq.(\ref{omge}).
The first three lines in Eq.(\ref{omgud}) correspond to the contribution
from the quarks taking part in the condensation while the 
fourth and fifth lines corrspond to the contribution
from  the two light quarks with the blue color.

Using Eq.s(\ref{omgud}) and (\ref{omgs}), the gap equations 
for the masses of u,d quarks are given by
\bearr
M_i-m_i&=&
\frac{8G_s}{(2\pi)^3}\int d\zbf k\frac{M_i}{\epsilon_i}
\Bigg(1+
\frac{\bar\xi_-}{\bar\omega_-}\left(\cos^2\theta_-^i-|\epsilon_{ij}|\
\sin^2\theta_-^j\right)
+
\frac{\bar\xi_+}{\bar\omega_+}\left(\cos^2\theta_+^i-|\epsilon_{ij}|\
\sin^2\theta_+^j\right)\nonumber\\
&-&\sin^2\theta_-^i-\sin^2\theta_+^i
+|\epsilon^{ij}|\left(\sin^2\theta_-j+\sin^2\theta_+^j\right)
-\sin^2\theta_-^{i3}-\sin^2\theta_+^{i3}\Bigg),
\label{mgapud}
\eearr
The mass gap equation for the strange quarks is given as
\be
M_3-m_3=\frac{8G_s}{(2\pi)^3}\int\frac{M_3}{\epsilon_3}\sum_{a=1,3}(1
-\sin^2\theta_-^{3a}-\sin^2\theta_+^{3a}).
\label{mgaps}
\ee

It is worthwhile to take the zero temperature limit of Eq.(\ref{omgud})
to compare with the results obtained earlier in Ref.\cite{igor,hmam}.
Using the relation
 $\lim_{\beta \rightarrow\infty}\frac{1}{\beta}
\ln(1+\exp(-\beta\omega))=-\omega\theta(-\omega)$, 
we have the contributions from the u,d quarks
beyond chiral symmetry restored case (i.e.$M_u=0=M_d$) given as
\bearr
\Omega_{ud}(T=0,M_{u,d}=0)&=&
\frac{3\Lambda^4}{2\pi^2}
-\sum_{i=1,2}\frac{\mu_{i3}^4}{12\pi^2}
\frac{2}{\pi^2}\int k^2\left(\sqrt{(k-\bar\mu)^2+\Delta^2}
+\sqrt{(k+\bar\mu)^2+\Delta^2}\right)dk\nonumber\\&+&
\frac{2}{\pi^2}\int_{\mu_-}^{\mu_+} k^2\left(\sqrt{\Delta^2+(k-\mu)^2}
+\delta_\mu\right)dk
\label{omgud0}
\eearr
where, $\mu_{\pm}=\bar\mu\pm\sqrt{\delta_\mu^2-\Delta^2}$. The limits of the
integrations over momentum in the third integral arises from the
conditions of nonnegativity of the argument of theta function for
nonzero contributions while taking the zero temperature limit.
It is reassuring that  the expression is identical as in 
Ref.\cite{igor}, where the two flavor colorsuperconductivity was
investigated.

Thus the thermodynamic potential is a function of three parameters: the two
mass gaps and a superconducting gap which need to be minimised
subjected to the conditions of electrical and color charge neutrality. 
The electric and charge neutrality constraints are given respectively as
\be
Q_E=\frac{2}{3}\rho^1-\frac{1}{3}\rho^2-\frac{1}{3}{\rho^3}
-\rho_e=0,
\label{qe}
\ee
and,
\be
Q_8 =\frac{1}{\sqrt{3}}\sum_ i (\rho^{i1} + \rho ^{i2}
- 2 \rho ^{i3}) =0.
\label{q8}
\ee
In the above $\rho^{ia}=\langle{\psi^{ia}}^\dagger\psi^{ia}\rangle
=2 I_v^{ia}$ ($i$, $a$ not summed) and $I_v^{ia}$ is as given in Eq.(\ref{iv}).
Further, $\rho^i=\sum_{a=1,3}\rho^{ia}$.
It may be worthwhile to note that the neutrality conditions Eq.(\ref{qe}) and
Eq.(\ref{q8}) can be combined to give rise to a simpler equation as
$3Q_E-\frac{\sqrt{3}}{2}Q_8=0$, i.e.
\be
\rho^{11}-\rho^{22}+\rho^{13}-\rho_e=0
\label{qe8}
\ee
At zero temperature, for the BCS phase, the number densities of
the condensing u and d quarks are the same i.e., $\rho^{11}=\rho^{22}$,
and Eq.(\ref{qe8}) can be solved for the electric chemical potential
as $\mu_E=-\frac{3}{5}(\mu-\frac{2}{\sqrt{3}}\mu_8)$.
It can also be shown by substituting this solution in Eq.(\ref{q8}) that
$\mu_8$ is very small compared to the chemical potentials $\mu_E$ and
$\mu$ as long as $\Delta/\mu$ is small \cite{igorr}.

On the other hand, for the case of gapless phase ($\Delta<|\delta_\mu|$),
one can solve Eq. (\ref{qe8}) for the gap at zero temperature as
\be
\Delta
=\sqrt{\frac{\mu_E^2}{4}-\left(\frac{\mu_{13}^3-\mu_E^3}{6\bar \mu^2}\right)}
\label{gapaprox}
\ee

\noindent Eq.(\ref{omgt}--\ref{q8}) and the superconducting 
gap equation Eq.(\ref{scgap})  constitute the basis of the numerical 
calculations that we shall discuss in the next section.

\section{Results and discussions}
\label{results}

For numerical calculations we have taken the values of
the parameters of NJL model as follows. The cutoff $\Lambda$
and the scalar coupling $G_s$ are chosen by fitting
the pion decay constant $f_\pi=93$ MeV and the chiral
condensate $\langle\bar u u\rangle^{1/3}=-250$ MeV=$\langle\bar d d
\rangle^{1/3}$. In case of non zero current quark masses, the
 additional parameters can be fixed from the values of pion and kaon masses.
In the following, however, we shall choose the current quark masses
for the light quarks as zero. This leads to the coupling
constand $G_s$ and the cut off $\Lambda$ as $G_s=5.0163$ GeV$^{-2}$ and
$\Lambda=0.6533$ GeV. Similar to Ref.\cite{bubnp} we  take the 
current quark mass of strange quarks
as $m_s=120$ MeV as a typical value giving ``reasonable" vacuum properties.
With this choice of parameters, the constituent quark 
masses at zero temperature and density are given as $M_1=0.313$ GeV=$M_2$,
and for strange quark $M_3=0.541$ GeV.  These values are similar to those
obtained in Ref.\cite{rehberg}, where the parameters have been fixed by
fitting vacuum masses and decay constants of pesudoscalar mesons.

Let us begin with the discussions of results 
without imposition of charge neutrality conditions. At zero temperature,
the behaviour of the gap parameters as a function of baryon chemical potential
(three times the quark chemical potential) is displayed in Fig.1-a.
We may point out here that these solutions for the gaps corresponds 
to the solutions for which the thermodynamic potential is minimised. In fact, 
in general, for certain values of the chemical potential there can be
several solutions of the gap equations, but we have
chosen the ones which minimise the thermodynamic potential. 
\begin{figure}
\vspace{-0.4cm}
\begin{center}
\begin{tabular}{c c }
\includegraphics[width=8cm,height=8cm]{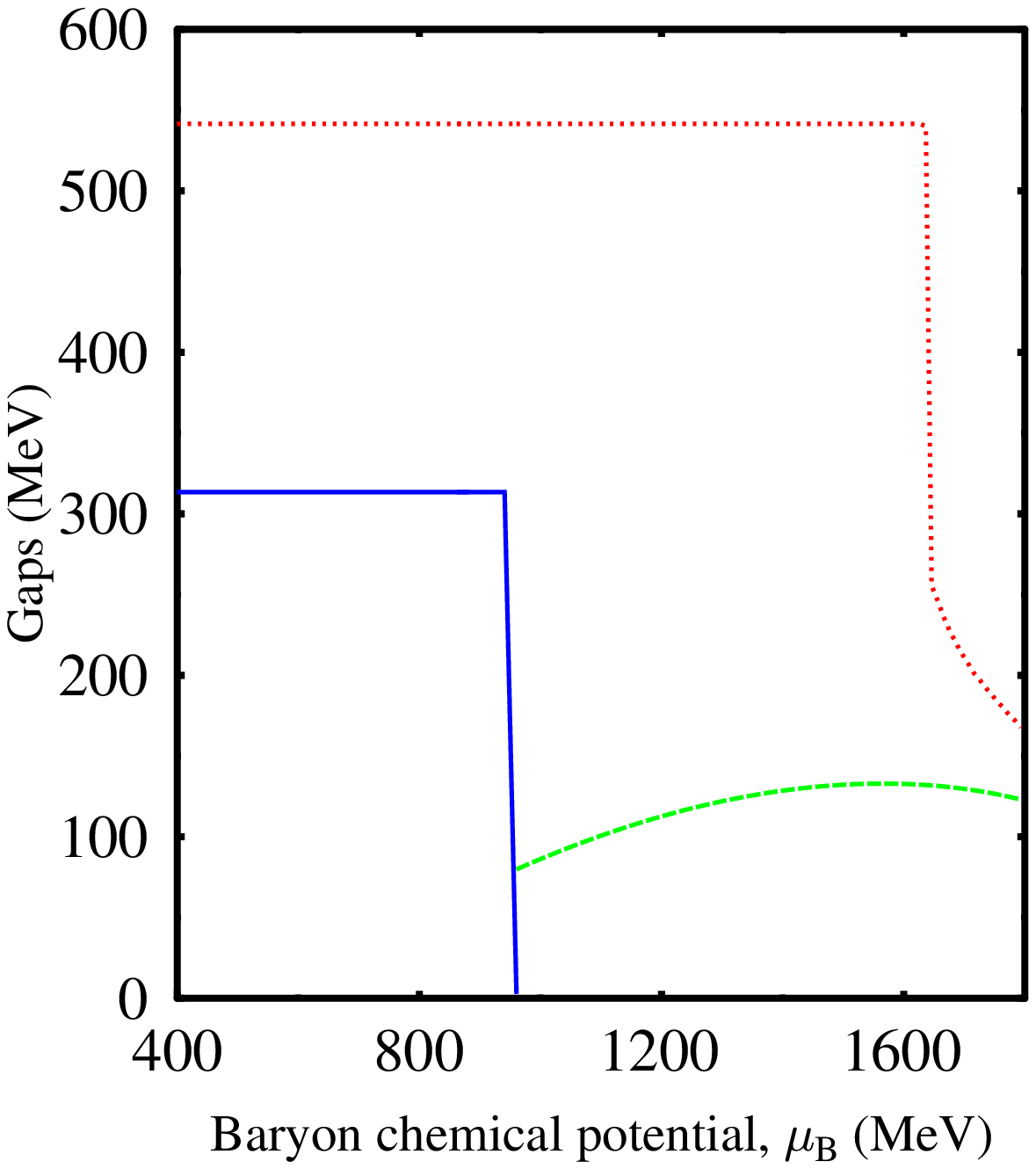}&
\includegraphics[width=8cm,height=8cm]{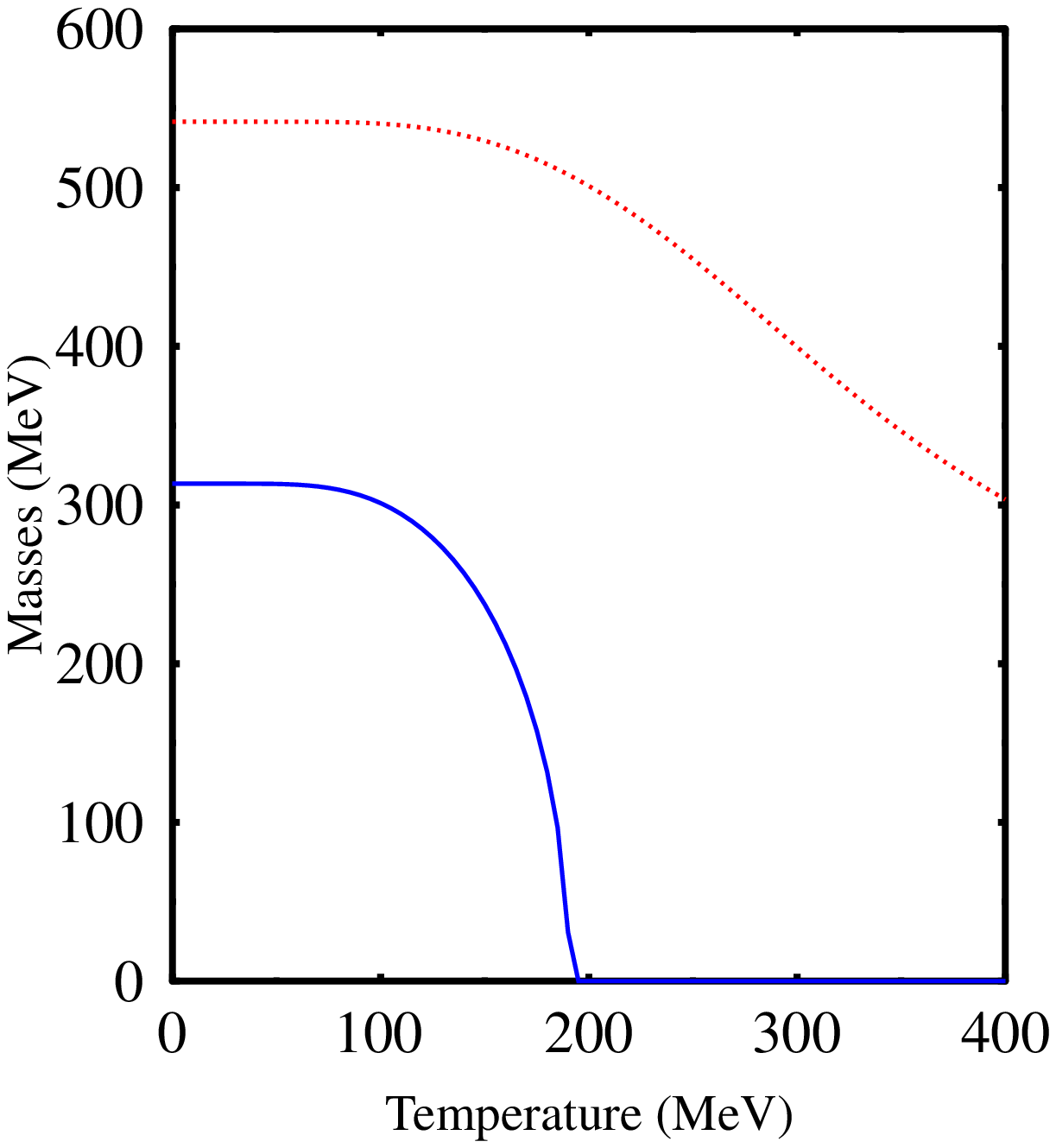}\\
Fig. 1-a & Fig.1-b
\end{tabular}
\end{center}
\caption{\it Gap parameters when charge neutrality conditions are not imposed.
Fig.1-a shows the gaps at zero temperature as a function of baryon chemical
potential. Fig. 1-b  shows the gaps at zero baryon density as a function of
temperature.
Solid curve refers to masses of u and d quarks, the dotted
curve refers to  mass of strange quarks and the dashed curve corresponds to 
the superconducting gap.}
\label{fig1}
\end{figure}

At low chemical potentials $\mu_B<\mu_1\sim$960 MeV, the diquark gap vanishes
and the masses of the quarks stay at their vacuum values. The entire region 
below $\mu_B=\mu_1$ corresponds to vacuum solution and has zero baryon number.
At $\mu_B=\mu_1$ a first order phase transition takes place and the
system is a two flavor color superconductor. The diquark gap jumps from zero to
about 80 MeV at this point. At the same point, the masses of u and d quarks 
drop from their vacuum values to zero. 

A first order transition is characterised by the existence of meta stable 
phases, the equivalent of e.g. oversaturated vapour. The masses 
corresponding to these metastable phases are the nontrivial ($M\neq 0$) 
solutions of the mass gap equation but have higher thermodynamic potential 
as compared to the solution corresponding to the stable phase.

With increasing $\mu_B$ the  superconducting gap increases until it
reaches a maximum at $\mu_B\sim 1565$ MeV. 
The baryon number density becomes nonzero
in the superconducting phase. At $\mu_B=\mu_1$, it jumps from zero to 
0.34 fm$^{-3}$ (around twice the nuclear matter density). The strange quark 
density remains zero till
$\mu_B\sim 1630$ MeV as strange quark mass remains at its vacuum value 
in this regime. Just beyond this point, the constituent mass of strange quark 
drops to about half its vacuum value and then decreases slowly.
The density of strange quarks also becomes nonzero beyond
$\mu_B=1630$ MeV. This will have important consequences regarding charge 
neutrality conditions as we shall discuss little later.
A flavor mixing interaction probably could lead to a reduced value of the strange
quark and hence to a lower threshold for nonzero density of strange quarks.
At larger chemical potential, the results become cutoff dependent and the 
diquark condensate decreases with $\mu_B$. We also note here that contrary 
to the vector-interactions of Ref.\cite{hmam}, we do not find a window 
in chemical potential where both chiral and the diquark condensate coexist.

In Fig. 1-b we show the temperature
dependence of masses at zero density. Chiral symmetry is restored for
light quarks at temperature about 195 MeV. The sharp first order transition
of zero temperature becomes a second order transition at 
zero baryon density as is reflected in the smooth
variation of the mass which is proportional to the order parameter
$\langle\bar\psi\psi\rangle$. 
\begin{figure}[htbp]
\begin{center}
\includegraphics[width=8cm,height=8cm]{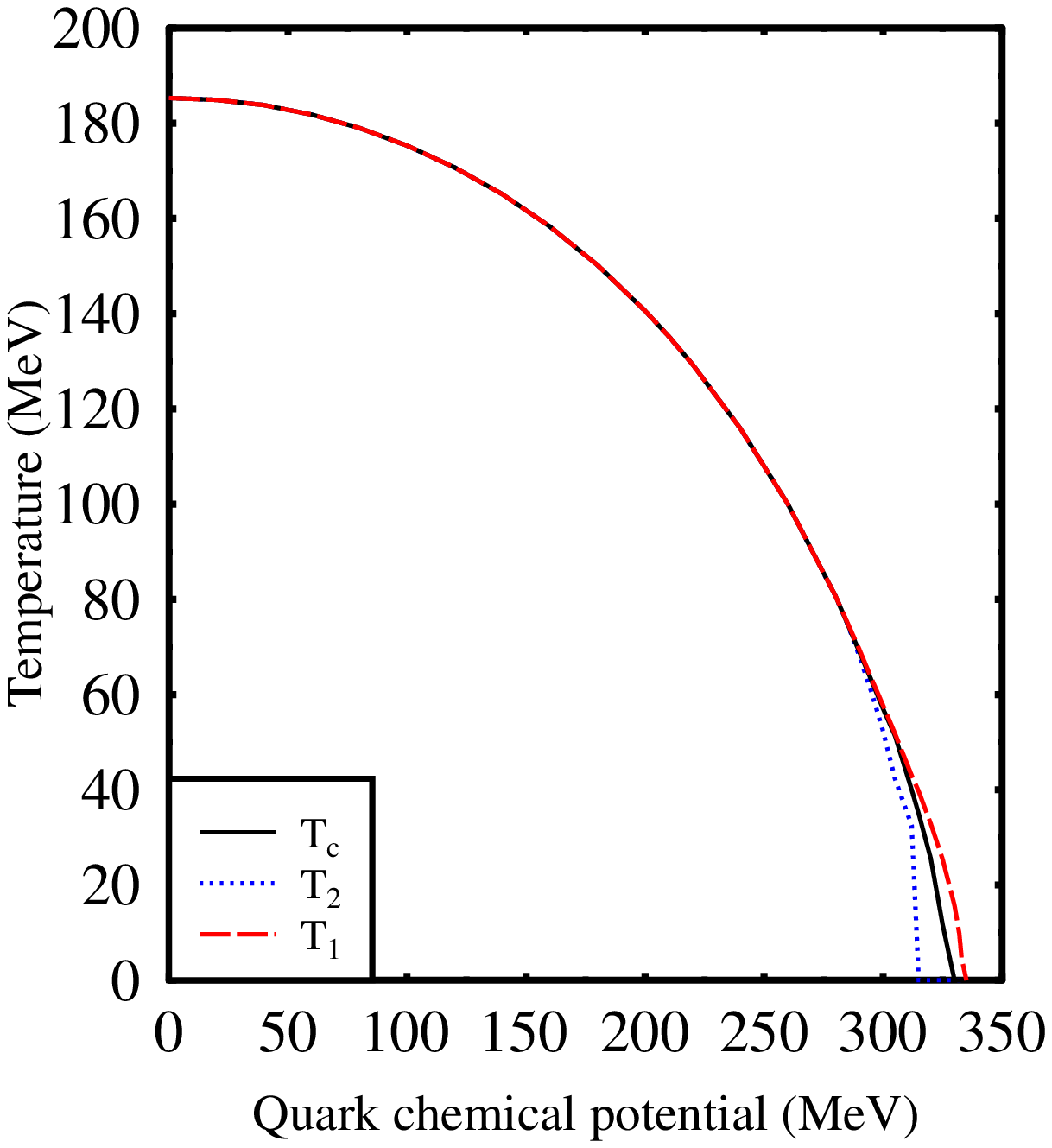}
\end{center}
\caption{\em Phase diagram in the ($\mu$,T) plane. Middle curve is the
critical line and the outer lines are the lower and upper spinodals}
\label{phasefig}
 \end{figure}

In Fig.\ref{phasefig}
 we show the resulting phase diagram for the chiral transition in
the temperature  and quark chemical potential ($1/3\mu_B$) 
plane in the present model when charge neutrality conditions are
not imposed. The middle line
is the critical line and corresponds to the state where the two nontrivial 
solution of the gap equations
have the same thermodynamic potential. Along this line the thermodynamic
potential has two minima of equal depth separated by a barrier and the
barrier height decreases with temperature. At the tricritical point
this barrier vanishes. Beyond this point there is only one solution for
the gap equation and the transition is second order. This critical
point ($T_c,\mu_c$) turns out to be (74.9,285 ) MeV. The other two lines 
are the spinodal lines constraining regions of spinodal instability.
We might note here that for this transition the strange quarks do not play
a role as their mass is much too large to contribute to the dynamics.

\begin{figure}[htbp]
\begin{center}
\includegraphics[width=8cm,height=8cm]{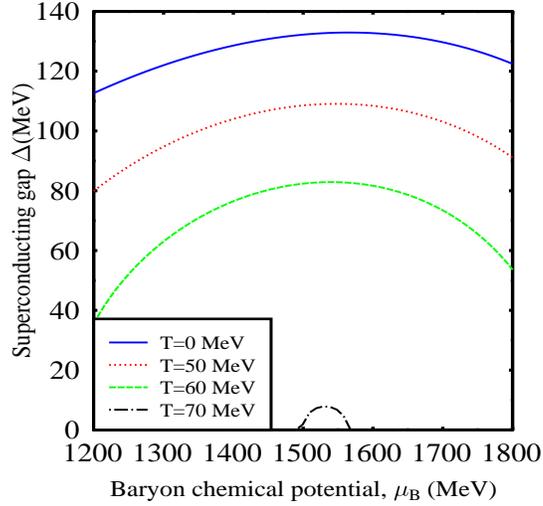}
\end{center}
\caption{\em Superconducting gap as a function of baryon density
for different temperatures. Here charge neutrality conditions 
(i.e.$\mu_E$=0=$\mu_8$) are not imposed.}
\label{gapfig}
 \end{figure}

In Fig.\ref{gapfig} we display the density dependence
of the superconducting gap for various temperatures.
 Similar to various BCS-type calculations,
this transition is of the second order.
The critical temperature is about half of the gap value at zero 
temperature as is in BCS case. This phase transition is a second
order phase transition  with the transition temperature
$T_c\simeq 0.57 \Delta(T=0)$.

\begin{figure}[htbp]
\begin{center}
\includegraphics[width=8cm,height=8cm]{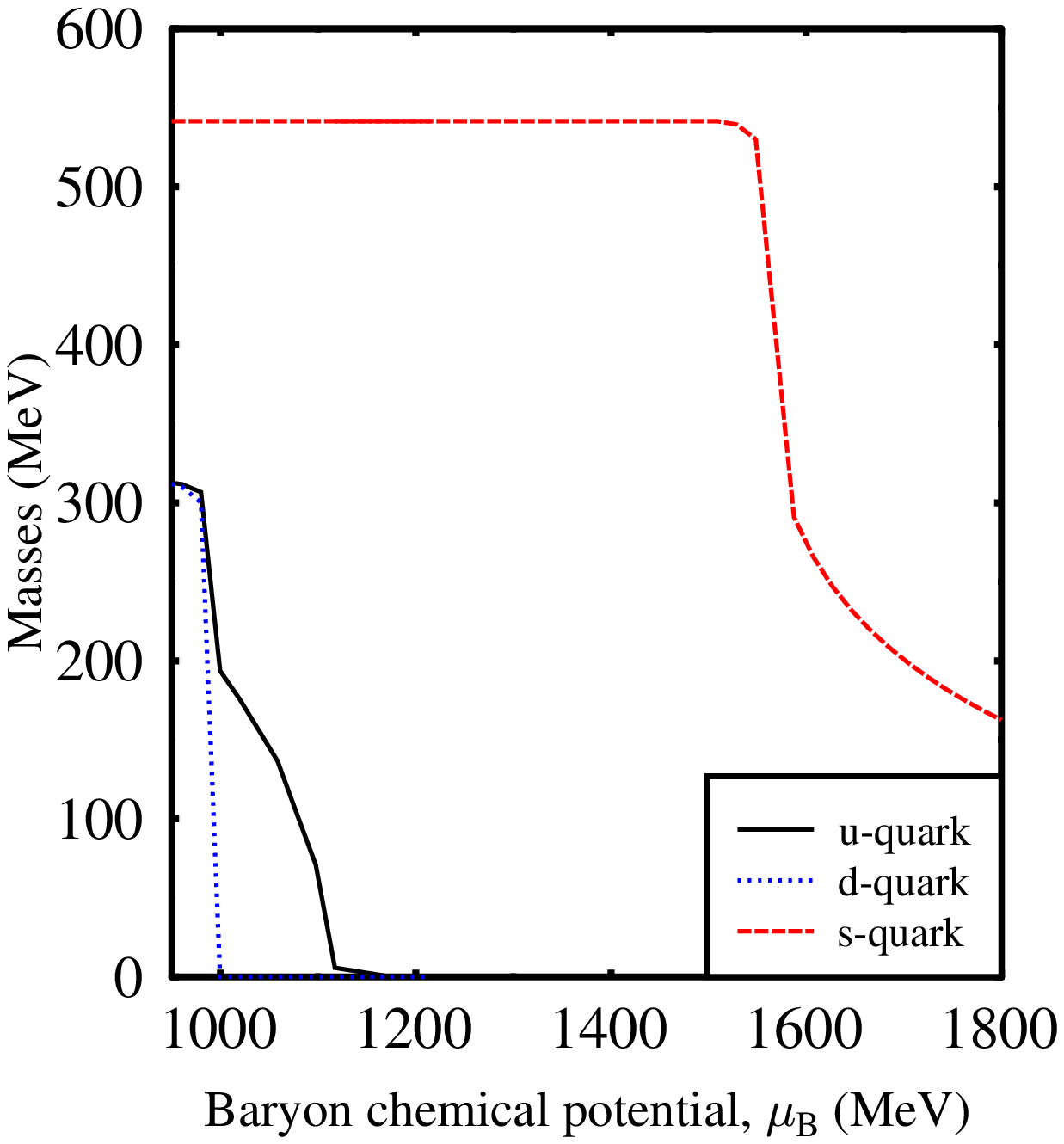}
\end{center}
\caption{\em{ Masses of u-quarks (solid line), d-quarks (dotted line) and
s-quarks (dashed line) as a function of baryon chemical potential when
charge neutrality conditions are imposed.}}
\label{masnjlchfig}
\end{figure}

We next extend our discussions to the case where the charge neutrality 
conditions ($\mu_E$=0=$\mu_8$) are imposed. In Fig.\ref{masnjlchfig} 
we show the dependence of masses when these conditions are imposed. 
For low baryon chemical potential, we see that the d quark masses 
start decreasing earlier than that of u- quark.
Charge neutrality conditions force the d quark density to be larger
(almost twice) than that of u-quark density making their masses 
vanishing earlier.

\begin{figure}
\vspace{-0.4cm}
\begin{center}
\begin{tabular}{c c }
\includegraphics[width=6.4cm,height=6.4cm]{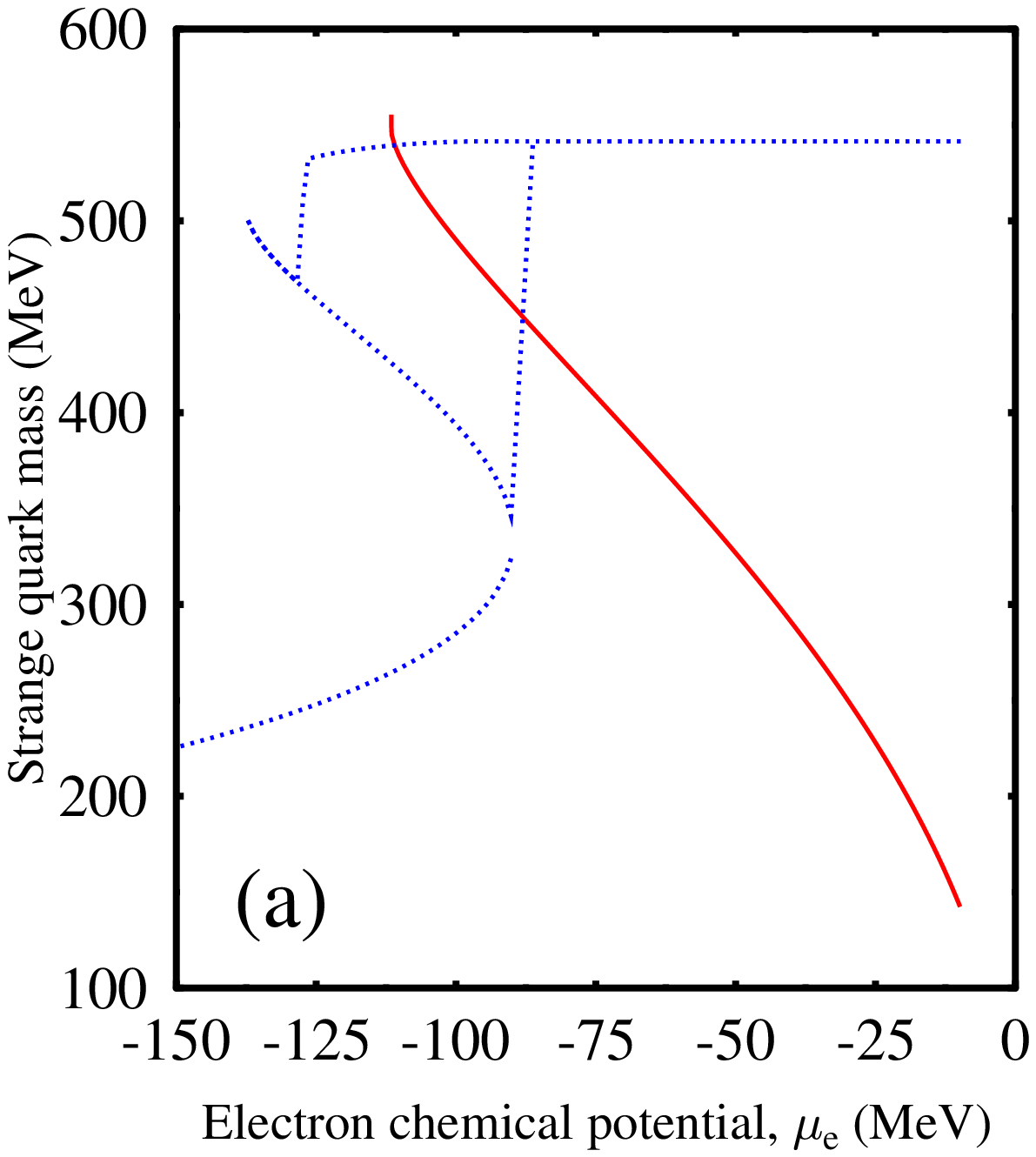}&
\includegraphics[width=6.4cm,height=6.4cm]{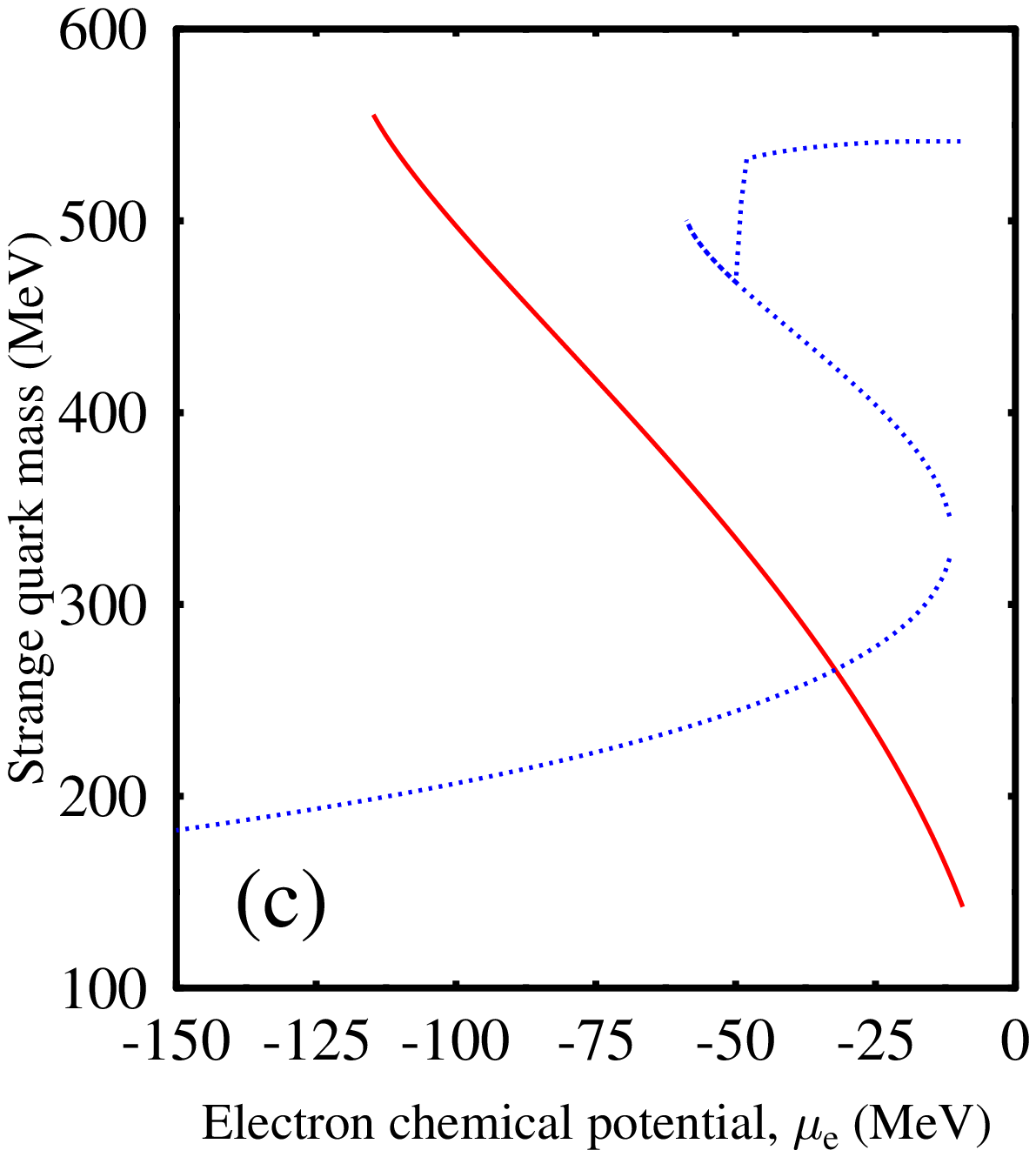} \\
Fig. 5-a & 
Fig.5-b 
\end{tabular}
\end{center}
\caption{\it 
Solutions of strange quark mass gap equation and charge neutrality condition 
at zero temperature.
In all the figures the electrical charge neutrality line is denoted
by the solid line and solutions of the mass gap equation for the strange quark 
is denoted by dotted line for baryon chemical
potential $\mu_B=1528$ MeV (Fig. 4-a) and
$\mu_B=1606$ MeV (Fig 4-b)}
\label{chstrangefig}
\end{figure}

At higher chemical potentials, strange quarks help in maintaining the charge
neutrality condition. We might note here that, in general there could be
several solutions of the mass gap solutions. A solution which may be 
free energetically favorable when chrage neutrality condition is not imposed
can ingeneral be disfavored when charge neutrality condition is imposed.
A typical example is shown in Fig.\ref{chstrangefig} for strange quark mass gap
for electrically neutral normal ($\Delta=0$) u-d-s matter.

In Fig. \ref{chstrangefig}, we  display graphical solutions for strange
quark mass when electrical charge neutrality condition is imposed.
The two figures -- Fig.5-a--b correspond to two different
values for chemical potentials $\mu_B$. Each value of electron 
chemical potential given in x axis then gives the value of the 
strange quark chemical potential
as $\mu_s=\mu_B-(1/3)\mu_E$. The dotted lines in these figures gives the
solutions of the mass gap equation for the strange quark for each $\mu_s$.
As may be seen there can be three 
different solutions for given baryon chemical potential and a given value 
of electric chemical potential.
We have also plotted here the charge neutrality line i.e. strange quark mass
parameter as a function of electric chemical potential such that the total
electric charge is zero. Thus, along the solid line the charge neutrality 
condition is satisfied. Along the dashed line mass gap equation is satisfied. 
The intersection of the charge neutrality line with the
curves satisfying the gap equation defines the desired solution satisfying 
both gap equation and the charge neutrality condition.
When charge neutrality condition is not imposed, the one which has minimum 
free energy is the required solution. However, this need not be the 
solution when
charge neutrality condition is imposed. Of the two cases, shown in 
Fig. \ref{chstrangefig}, for $\mu_B=1528$ MeV (Fig. 5-a), the solution with
masss about 520 MeV is the one which satisfies the charge neutrality condition.
For $\mu_B=1606$ MeV, (Fig. 5-b) $M_s\simeq 266$ MeV satisfies the 
charge neutrality condition. We might note here that, 
when charge neutrality condition is not imposed, the branch with
$M_s\simeq 520$ MeV is free energetically preferred solution
for baryon chemical potential $\mu_B$ as large as
1630 MeV as we mentioned earlier. The other two solutions although
free energetically unstable when charge neutrality condition is not imposed
one of them becomes the preferred solution when charge neutrality condition
is imposed. Similar thing happens when we impose charge neutrality conditions
for the superconducting gap as we shall see shortly.
\begin{figure}
\vspace{-0.4cm}
\begin{center}
\begin{tabular}{c c }
\includegraphics[width=8cm,height=8cm]{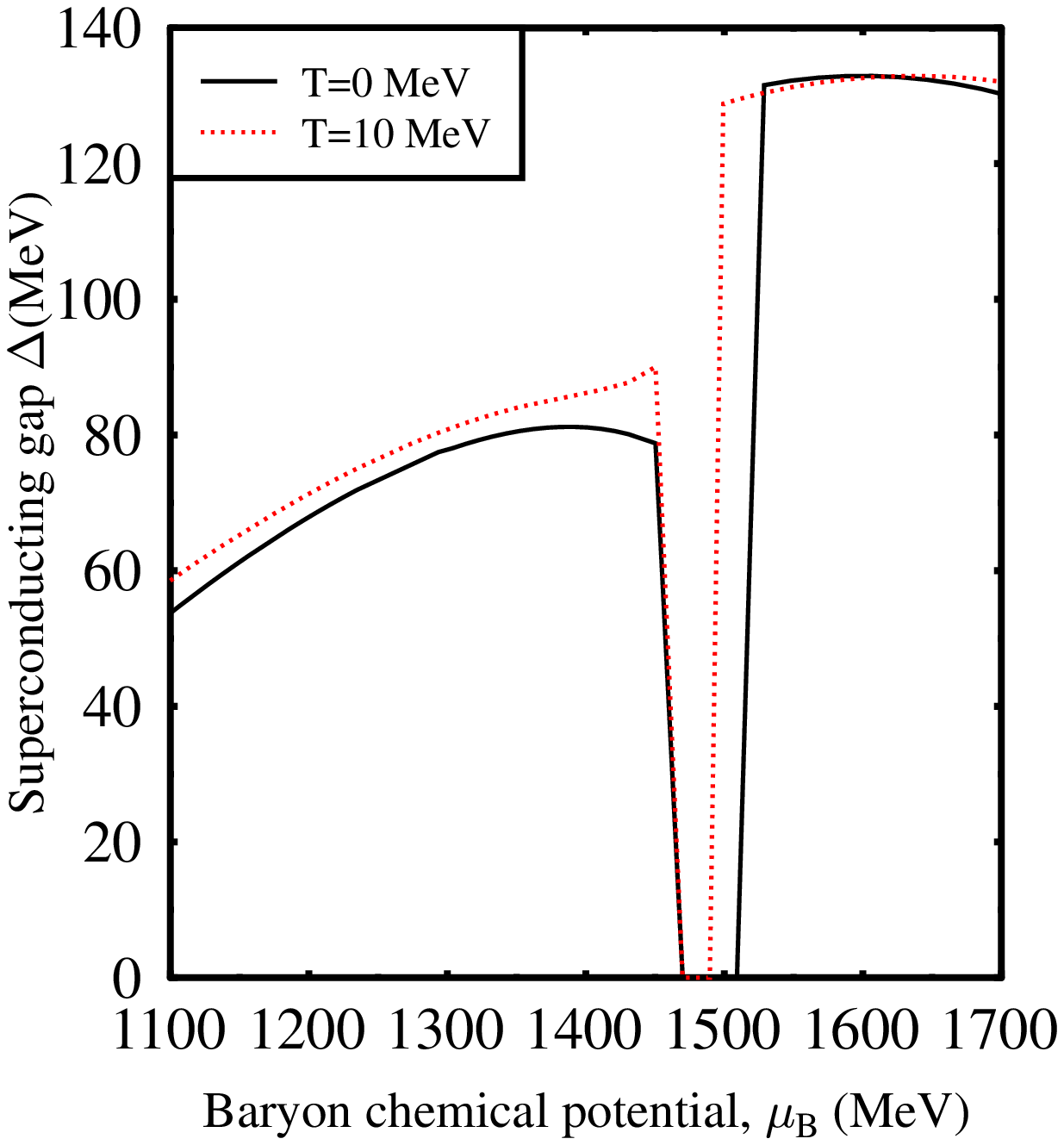}&
\includegraphics[width=8cm,height=8cm]{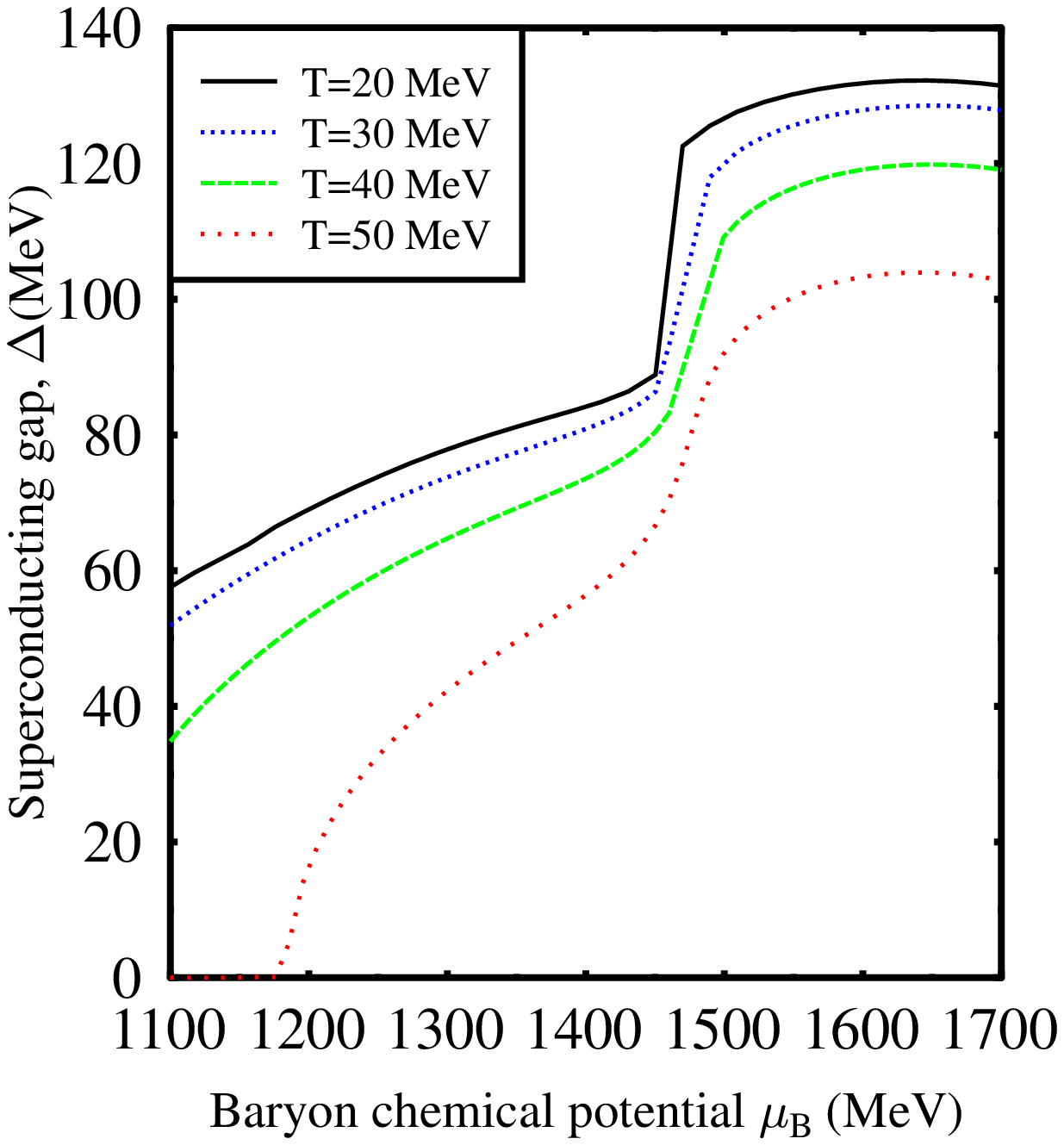}\\
Fig. 6-a & Fig.6-b
\end{tabular}
\end{center}
\caption{\it Superconducting gap at different temperatures
as a function of the baryon chemical potential,
when the color and electrical charge neutrality
conditions are imposed. 
}
\label{figdeln}
\end{figure}
\begin{figure}[htbp]
\begin{center}
 \includegraphics[width=10cm,height=12cm]{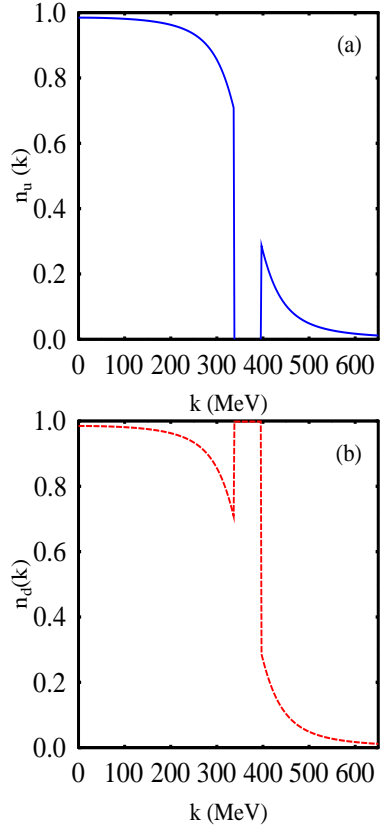}
\end{center}
\caption{\em Occupation number as a function of momentum. Solid line corresponds to u quarks and dotted line corresponds to d quarks}
\label{knkfig}
 \end{figure}
\begin{figure}
\vspace{-0.4cm}
\begin{center}
\begin{tabular}{c c }
\includegraphics[width=8cm,height=8cm]{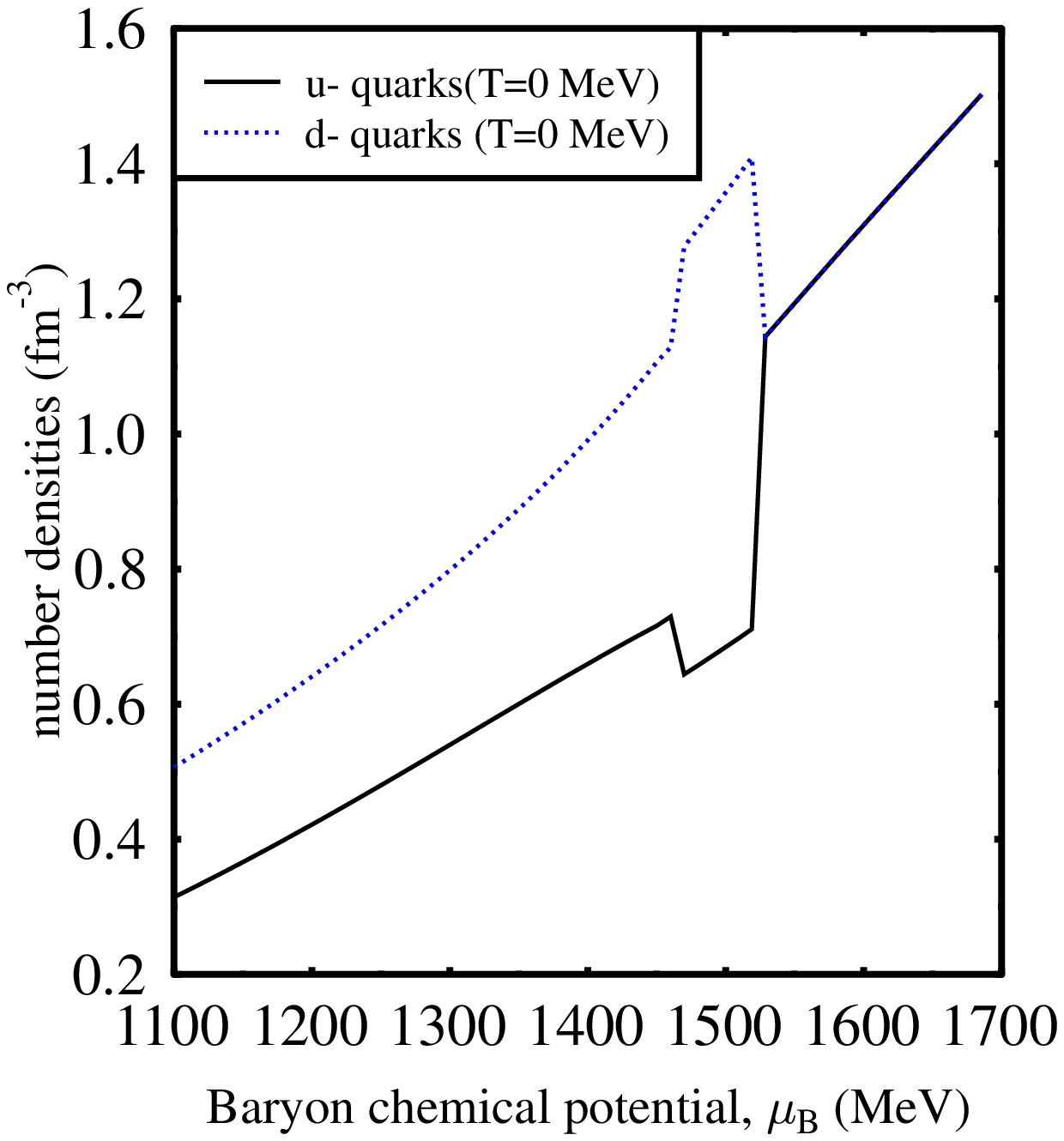}&
\includegraphics[width=8cm,height=8cm]{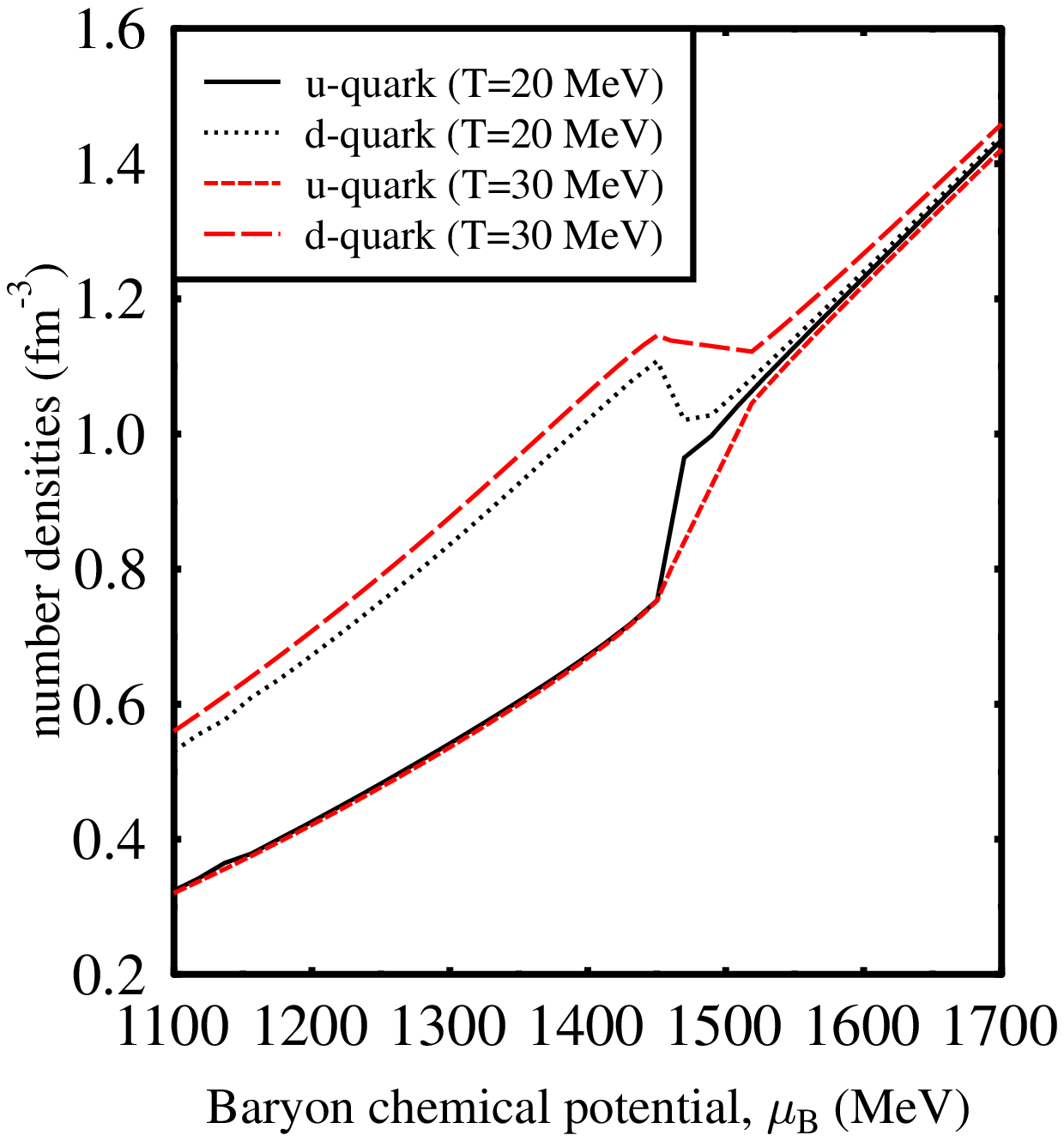}\\
Fig. 8-a & Fig.8-b
\end{tabular}
\caption{\em Number densities  of u quarks (solid) and d quarks (dot-dashed) 
participating in superconducting phase at zero (Fig 8.a) and at finite
 temperatures (fig. 8.b).}
\label{figdensc}
\end{center}
\end{figure}

\begin{figure}[htbp]
\begin{center}
\includegraphics[width=8cm,height=8cm]{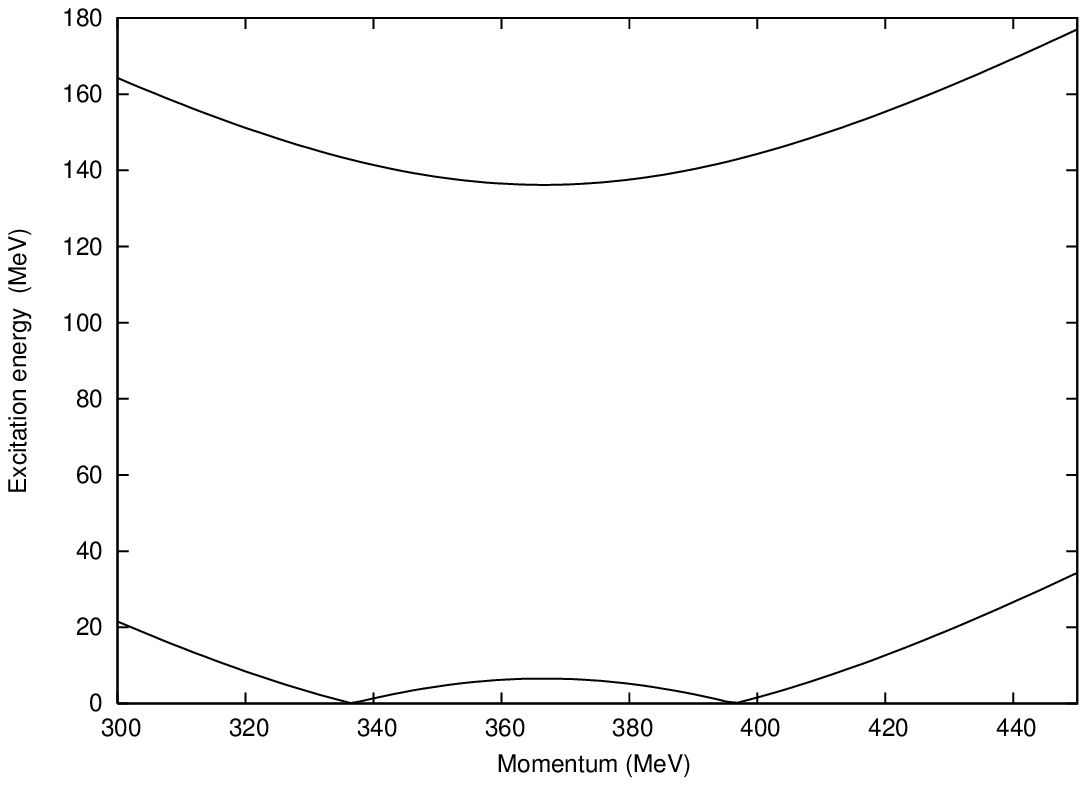}
\end{center}
\caption{\em Excitation energies as a function of chemical
potential.}
\label{dispfig}
 \end{figure}

We have also verified that in presence of both electric and color
charge neutrality conditions, we do not have a simultaneous diquark and 
quark antiquark condensation. 

 We compute the thermodynamic potential numerically as follows.
For a given temperature and quark chemical
potential, the thermodynamic potential is minimised with respect to the
strange quark constituent mass after solving the superconducting 
gap equation Eq.(\ref{scgap}) selfconsistently. The values of $\mu_E$ 
and $\mu_8$ are varied so that the charge neutrality 
conditions Eq.(\ref{qe}) and Eq.(\ref{q8}) are satisfied. The resulting 
superconducting gap is shown in Fig.\ref{figdeln}
for different temperatures.

As may be seen for smaller values of the baryon chemical potential
($\mu_B < 1470$ MeV) we have smaller values of the gap which reaches a
maximum about 80 MeV at $\mu_B=1390$ MeV at zero temperature.
Then it decreases and  vanishes  at $\mu_B=1470$ MeV=$\mu_2$ . We have also
verified that the difference between  the value of the gap  as obtained 
numerically satisfying the charge neutrality condition and the
the approximate expression given in Eq.(\ref{gapaprox}) equation is less than
a percent for the gapless phase. The gap remains zero till 
$\mu_B=\mu_3=1530$ MeV. In fact, in the region between
$\mu_2$ and $\mu_3$ we do not have any real solution for
$\Delta$. At $\mu_3$ the system jumps from the normal phase
to the BCS phase with a gap about 132 MeV. The baryon number density jumps from 
$\rho_B=0.96$fm$^{-3}$ to 1.54 fm$^{-3}$ i.e. from 6 to 9.5 times
the nuclear matter density. The interval in the  baryon chemical potential
between the  gapless and BCS phase within which it is normal quark matter
decreases with increase in temperature and then disappears as may be noted 
in Fig.s\ref{figdeln}. The sharp transition between gapless phase to
the BCS phase  as a function of baryon chemical potential 
at smaller temperature becomes a smooth transition
as temperature increases as may be seen from Fig. 6--b. We also would like
to note that the starnge quark number density at zero temperature 
is zero in the gapless phase. This is in contrary to observation made in
Ref.\cite{hmam} where a vector vector point interaction was considered.
However, as temperature is increased, the number densitiy of the strange 
quarks become nonzero. E.g., at a temperature of 50 MeV, strange quarks 
contribute upto about 15 $\%$ of that of density of u--quarks.

Let us next discuss now some of the characteristics of the gapless modes which
occur for smaller values of the baryon chemical potential ($\mu_B<\mu_2$).
In fact, in this region, in general, there are two solutions to gap 
equations for given $\bar \mu$ and $\delta_\mu$. 
One is the usual BCS solution with a larger
value of the gap and the other one with a smaller value of the gap.
However, it so happens that in this range of the chemical potential, the 
solution of the gap equation with larger value of the gap (the BCS solution)
does not satisfy the charge neutrality condition. The situation here is
similar to the case for strange quark mass when charge neutrality 
condition is imposed.

At zero temperature, gapless  modes occur  when the gap is less 
than half the difference of the chemical potential $\delta\mu$ 
of the two condensing quarks. It is easy to
show also that the excitation
energy $\omega_2$ of the d-quark vanishes at momenta 
$\mu_- = \bar\mu-\sqrt{\delta\mu^2-\Delta^2}$ and 
$\mu_+= \bar\mu+\sqrt{\delta\mu^2-\Delta^2}$ at zero temperature.
In this phase the number densities of the condensing u and d quarks are
not the same. Let us note that the number densities of u-red quarks 
are given as
\be
\rho^{11}=\langle{\psi^{11}}^\dagger\psi^{11}\rangle=\frac{2}{(2\pi)^3}
\int d\zbf k(F^{11}-F_1^{11})=\rho^{12}
\ee
where
\be
F^{11}=\sin^2 \theta_-^{1}-\frac{1}{2}\left(1-\frac{\bar\xi_-}{\bar\omega_-}\right)
\left(1-\sin^2\theta_-^1-\sin^2\theta_-^2\right)
\label{f11}
\ee
for the quarks and 
\be
F_1^{11}=\sin^2 \theta_+^{1}-\frac{1}{2}\left(1-\frac{\bar\xi_+}{\bar\omega_+}\right)
\left(1-\sin^2\theta_+^1-\sin^2\theta_+^2\right)
\label{f22}
\ee
for the antiquarks. Interchanging indices $1$ and $2$ we shall have the
expression for the number density of the d quarks taking part in
condensation. At zero temperature, the distribution function 
$\sin^2\theta_-^1$ vanishes for the u quarks as coresponding energy
$\omega_1=\bar\omega-\delta_\mu$ is always positive 
(let us remember that $\delta\mu$ is negative at zero temperature) 
 where as the distribution function
for d quarks is nonzero between $\mu_-$ and $\mu_+$. In fact, 
the number densities at zero temperature are given as
\be
\rho^u=2\rho^{11}=\frac{4}{(2\pi)^3}\int
d\zbf k\frac{1}{2}\left[\left(1-\frac{\bar\xi_-}{\bar\omega_-}\right)
\left(1-\theta(-\omega_2)\right)-\left(1-
\frac{\bar\xi_+}{\bar\omega_+}\right)\right]
\label{rhou}
\ee
for  the u quarks. Similarly the density of the d quarks taking part 
in condensation is given at zero temperature by
\be
\rho^d=2\rho^{21}=\frac{4}{(2\pi)^3}\int
d\zbf k\left[\theta(-\omega_2)+\frac{1}{2}\left(1-\frac{\bar\xi_-}{\bar\omega_-}
\right)
\left(1-\theta(-\omega_2)\right)-\left(1-
\frac{\bar\xi_+}{\bar\omega_+}\right)\right]
\label{rhod}
\ee
The integrands of Eq.s (\ref{rhou}) and (\ref{rhod}) which are the 
occupation numbers in the momentum space are plotted in Fig.\ref{knkfig}. 
It is easy to see from the expressions in Eq.s(\ref{rhou}) and 
(\ref{rhod}) that except in the interval $(\mu_-,\mu_+)$, 
in the momentum space, the theta function does not contribute and
the distribution is like the BCS distribution. The occupation number 
distribution is a `dented' theta function with the `dent' $\sim \Delta$
around the average fermi surface.
In the region  between $\mu_-$ and 
$\mu_+$, the occupation numbers resemble like that of normal matter as
$\rho_u$ becomes almost zero but for the  vanishingly small 
antiparticle contribution. On the other hand, the  integrand of $\rho_d$ 
becomes unity in the same limit for the antiparticle contributions.
 This is shown in Fig.\ref{knkfig}. It follows therefore that the total
number of u-quarks and d-quarks (integrated over the momentum) are
not the same. The difference in their number densities is in fact given by
\be
\delta\rho_{sc}=\rho_2-\rho_1=\frac{4}{(2\pi)^3}
\int d\zbf k\left[\left(\sin^2\theta_-^2-\sin^2\theta_-^1\right)+
\left(\sin^2\theta_+^2-\sin^2\theta_+^1\right)\right].
\label{drho}
\ee
Thus at finite temperatures the $\delta\rho_{sc}$ will always be nonzero
as long as the difference in the chemical potential is nonzero. At zero
temperature this difference will be nonzero only for the gapless solution 
i.e.  for gaps less than half the magnitude of the difference
in chemical potentials. This is infact given by 
$\delta\rho_{sc}(T=0)=2/(3\pi^2)(\mu_+^3-\mu_-^3)$.
In Fig.\ref{figdensc}, we plot the number densities of
u and d quarks taking part in the condensation. At zero temperature, 
the number densities become the same in the BCS phase. However, because of 
thermal distributions, the two densities do not become exactly 
identical even in the BCS phase at finite temperatures.

 In Fig.\ref{dispfig} we plot the dispersion relations - the excitation
energies  of the quasi particles as a function of momentum for 
quark chemical potential $\mu_q=(1/3)\mu_B\simeq 379$MeV and at zero 
temperature.  Let us note 
that the dispersion relations are given by $\omega_1=\bar\omega-\delta_\mu$
 and $\omega_2= \bar\omega+\delta_\mu$. The gap in this case turns out to be
$\Delta=59$ MeV and the difference of the chemical potential
turns out to be $\delta_\mu\simeq -66$ MeV at zero temperature. Far
from the pairing region $\bar\mu\simeq 354$ MeV, the spectrum looks 
like that of usual BCS type dispersion relations. Of the two excitation 
energies, $\omega_1$ shows a minimum at the average fermi momentum $\bar\mu$
 with a value $\omega_1(|\zbf k|=\bar\mu)=\Delta+\delta_\mu\simeq 125$MeV. 
On the other hand, $\omega_2$ becomes gapless ($\omega_2=0$)at
momenta $\mu_-$ and $\mu_+$. In this `breached' pairing
region \cite{wilczek} we have only unpaired down quarks and no up quarks.

\begin{figure}[htbp]
\begin{center}
 \includegraphics[width=8cm,height=8cm]{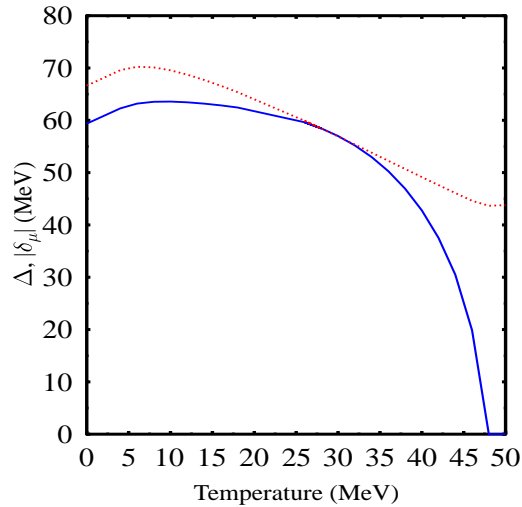}
\end{center}
\caption{\em Temperature dependence of gap parameter in the gapless
phase (solid line). $|\delta_\mu|$ as a function of temperature is shown as the 
dotted line.}
\label{gapltfig}
 \end{figure}

In Fig.\ref{gapltfig} we show the temperature dependence of the
gapless modes. We have taken the quark chemical potential as 
379 MeV as earlier. Initially the gap increases and then
decreases monotonically and vanishes at $T_C=48$MeV i.e. 
$T_C=0.81 \Delta(T=0)$. Clearly, this shows a departure from the BCS 
behaviour with  $T_C/\Delta(T=0)\neq 0.567$. Further, it is also seen
that for temperatures between 28 MeV to 32 MeV, the gap is
larger than difference of chemical potential and $\omega_2$ no longer is
exactly gapless in between these temperatures. The excitation energies however
are much smaller than temperature and are almost similar to the excitation 
energies between $\mu_-$ and $\mu_+$ at zero temperature.
For temperatures beyond 32 MeV,
the gap becomes smaller than $|\delta_\mu|$ and we 
have again the gapless modes occuring at corresponding $\mu_-$ and $\mu_+$.

It may be worthwhile to mention here that gapless modes in
superconductivity were known theoretically in the context of 
superconducting matter with finite momentum \cite{abrikosov} 
as well as in condensed matter systems with magnetic impurities 
\cite{sarma}. Recently it has been 
investigated for cold fermionic atoms \cite{wilczek,rapid}. 
Gapless modes in the
context of quark matter has been first proposed in Ref.\cite{krischprl} for
color flavor locked matter. However, this corresponded to a meta 
stable phase. Gapless modes in neutral quark matter were first emphasized
for the two flavor colorsuperconductivity in Ref. \cite{igor}
and for the 2SC+s quark matter in Ref. \cite{hmam}.
Stable gapless modes for color flavor
locked matter has been first proposed in Ref.\cite{kriscfl} and has 
been confirmed in Ref.  \cite{shovris} in a more general structure
for the gap. The temperature dependence of the CFL gapless modes
has been studied in Ref.\cite{krisaug}.



As the chemical potential is increased beyond 1530 MeV at zero temperature
the solution for the gap jumps to a higher value of
about 130 MeV almost similar to the case when charge neutrality
condition is not imposed. This corresponds to the usual BCS
solution. In this case, 
the numbers of u quarks and d quarks participating in condensation
are the same. The charge neutrality condition however is maintained 
by the third color, the electron as well as the strange quarks.
One essential effect of including strange quarks is that the electron density
starts decreasing for chemical potentials greater than the strange quark
mass as the strange quark can carry the negative charge to maintain
electric charge neutrality condition. This has the effect that
the lowest excitation energy $\omega_2=\bar\omega+\delta_\mu$
in the BCS pairing case becomes large due to both the large
value of the gap as well as the smaller magnitude of the
electron chemical potential.

\section{Summary}
\label{summary}
                 We have analysed here in NJL model
the structure of vacuum in terms of quark--antiquark 
as well as diquark pairing at finite temperature. The methodology 
uses an explicit variational construct of the trial state.
The gap function as well as the thermal  distribution functions are 
also determined variationally. 
To consider neutron star matter, we have imposed the
color and electric charge neutrality conditions through 
the introduction of appropriate chemical potentials. It has been noted 
and emphasized earlier that projecting out the color singlet 
state from color neutral state costs 
negligible free energy for large enough chunk of color neutral matter
\cite{krisch}.

We observe that the mass of the strange quark plays a sensitive role in 
maintaining charge neutrality condition.  Further, we observed that solutions
for the gap quations which may not be free energtically preferble solutions,
can be the preferred solutions when constraints of charge neutrality 
conditions are imposed.  We also find that with increase in chemical potential,
the quark matter has a transition from a gapless phase to the BCS phase 
through an intermediate normal quark matter phase. This interval between the
gapless and BCS phase decreases and then disappears with increase in the
temperature. We also find that there are no strange quarks in the gapless
phase at zero temperaure contrary to the observation in the vectorial point
 interaction model as in Ref.\cite{hmam}. However, at higher temperatures only
the contribution of strange quarks becomes significant even in the gapless mode.
We have looked into the gapless  to 
BCS transition as a function of density for various temperatures.
The  sharp transition at zero temperature
becomes a smooth transition at higher temperatures as density is increased.

We have focussed our attention here to the  two flavor superconducting phase
with strange quarks. 
The variational method adopted can be directly generalised 
to include color flavor locked 
phase and one can then make a free energy comparison
regarding the possibility of which phase would be thermodynamicaly favourable
at what density. This will be particularly interesting for
cooling of neutron stars with a CFL core.
We have considered here homogeneous phase of matter. 
However, we can also consider mixed phases of matter with matter being neutral
on the average. Surface tension between the suerconducting phase and the
normal quark phase will become an important factor in determining
the stability of this mixed phase \cite{rupak}.
Some of these problems are being investigated and will be 
reported elsewhere \cite{hmampp}.

\acknowledgments
The authors would like to thank D.H. Rischke, I. Shovkovy, Mei Huang,
J.C. Parikh, H.C. Ren, P. Jaikumar and T. Kunihiro for many useful discussions.
One of the authors (HM) would like to thank Institute for
Nuclear Theory, Seattle for hospitality during the program INT-04/1
during the initial stages of the present investigation.
AM would like to thank Institut f\"ur Theorestische for warm hospitality
where the present work was initiated.

\def \ltg{R.P. Feynman, Nucl. Phys. B 188, 479 (1981); 
K.G. Wilson, Phys. Rev. \zbf  D10, 2445 (1974); J.B. Kogut,
Rev. Mod. Phys. \zbf  51, 659 (1979); ibid  \zbf 55, 775 (1983);
M. Creutz, Phys. Rev. Lett. 45, 313 (1980); ibid Phys. Rev. D21, 2308
(1980); T. Celik, J. Engels and H. Satz, Phys. Lett. B129, 323 (1983)}

\def\berges {J. Berges, K. Rajagopal, {\NPB{538}{215}{1999}}.}
\def \svz {M.A. Shifman, A.I. Vainshtein and V.I. Zakharov,
Nucl. Phys. B147, 385, 448 and 519 (1979);
R.A. Bertlmann, Acta Physica Austriaca 53, 305 (1981)}

\def \spmbst {S.P. Misra, Phys. Rev. D35, 2607 (1987)}

\def \hmgrnv { H. Mishra, S.P. Misra and A. Mishra,
Int. J. Mod. Phys. A3, 2331 (1988)}

\def \snss {A. Mishra, H. Mishra, S.P. Misra
and S.N. Nayak, Phys. Lett 251B, 541 (1990)}

\def \amqcd { A. Mishra, H. Mishra, S.P. Misra and S.N. Nayak,
Pramana (J. of Phys.) 37, 59 (1991). }
\def\qcdtb{A. Mishra, H. Mishra, S.P. Misra 
and S.N. Nayak, Z.  Phys. C 57, 233 (1993); A. Mishra, H. Mishra
and S.P. Misra, Z. Phys. C 58, 405 (1993)}

\def \spmtlk {S.P. Misra, Talk on {\it `Phase transitions in quantum field
theory'} in the Symposium on Statistical Mechanics and Quantum field theory, 
Calcutta, January, 1992, hep-ph/9212287}

\def \hmnj {H. Mishra and S.P. Misra, 
{\PRD{48}{5376}{1993}.}}

\def \hmqcd {A. Mishra, H. Mishra, V. Sheel, S.P. Misra and P.K. Panda,
hep-ph/9404255 (1994)}

\def \amcrl {A. Mishra, H. Mishra and S.P. Misra, Z. Phys. C 57, 241 (1993)}

\def \higgs { S.P. Misra, in {\it Phenomenology in Standard Model and Beyond}, 
Proceedings of the Workshop on High Energy Physics Phenomenology, Bombay,
edited by D.P. Roy and P. Roy (World Scientific, Singapore, 1989), p.346;
A. Mishra, H. Mishra, S.P. Misra and S.N. Nayak, Phys. Rev. D44, 110 (1991)}

\def \nmtr {A. Mishra, 
H. Mishra and S.P. Misra, Int. J. Mod. Phys. A5, 3391 (1990); H. Mishra,
 S.P. Misra, P.K. Panda and B.K. Parida, Int. J. Mod. Phys. E 1, 405, (1992);
 {\it ibid}, E 2, 547 (1993); A. Mishra, P.K. Panda, S. Schrum, J. Reinhardt
and W. Greiner, to appear in Phys. Rev. C}

\def \dtrn {P.K. Panda, R. Sahu and S.P. Misra, 
Phys. Rev C45, 2079 (1992)}

\def \qcd {G. K. Savvidy, Phys. Lett. 71B, 133 (1977);
S. G. Matinyan and G. K. Savvidy, Nucl. Phys. B134, 539 (1978); N. K. Nielsen
and P. Olesen, Nucl.  Phys. B144, 376 (1978); T. H. Hansson, K. Johnson,
C. Peterson Phys. Rev. D26, 2069 (1982)}

\def \cornwal {J.M. Cornwall, Phys. Rev. D26, 1453 (1982)}
\def\aichlin {F. Gastineau, R. Nebauer and J. Aichelin,
{\PRC{65}{045204}{2002}}.}

\def \mndglv {J. E. Mandula and M. Ogilvie, Phys. Lett. 185B, 127 (1987)}

\def \schwinger {J. Schwinger, Phys. Rev. 125, 1043 (1962); ibid,
127, 324 (1962)}

\def \schutte {D. Schutte, Phys. Rev. D31, 810 (1985)}

\def \amspm {A. Mishra and S.P. Misra, 
{\ZPC{58}{325}{1993}}.}

\def \gft{ For gauge fields in general, see e.g. E.S. Abers and 
B.W. Lee, Phys. Rep. 9C, 1 (1973)}

\def \gribov {V.N. Gribov, Nucl. Phys. B139, 1 (1978)}

\def \spm78 {S.P. Misra, Phys. Rev. D18, 1661 (1978); {\it ibid}
D18, 1673 (1978)} 

\def \lopr {A. Le Youanc, L.  Oliver, S. Ono, O. Pene and J.C. Raynal, 
Phys. Rev. Lett. 54, 506 (1985)}

\def \spphi {S.P. Misra and S. Panda, Pramana (J. Phys.) 27, 523 (1986);
S.P. Misra, {\it Proceedings of the Second Asia-Pacific Physics Conference},
edited by S. Chandrasekhar (World Scientfic, 1987) p. 369}

\def\spmdif {S.P. Misra and L. Maharana, Phys. Rev. D18, 4103 (1978); 
    S.P. Misra, A.R. Panda and B.K. Parida, Phys. Rev. Lett. 45, 322 (1980);
    S.P. Misra, A.R. Panda and B.K. Parida, Phys. Rev. D22, 1574 (1980)}

\def \spmvdm {S.P. Misra and L. Maharana, Phys. Rev. D18, 4018 (1978);
     S.P. Misra, L. Maharana and A.R. Panda, Phys. Rev. D22, 2744 (1980);
     L. Maharana,  S.P. Misra and A.R. Panda, Phys. Rev. D26, 1175 (1982)}

\def\spmthr {K. Biswal and S.P. Misra, Phys. Rev. D26, 3020 (1982);
               S.P. Misra, Phys. Rev. D28, 1169 (1983)}

\def \spmstr { S.P. Misra, Phys. Rev. D21, 1231 (1980)} 

\def \spmjet {S.P. Misra, A.R. Panda and B.K. Parida, Phys. Rev Lett. 
45, 322 (1980); S.P. Misra and A.R. Panda, Phys. Rev. D21, 3094 (1980);
  S.P. Misra, A.R. Panda and B.K. Parida, Phys. Rev. D23, 742 (1981);
  {\it ibid} D25, 2925 (1982)}

\def \arpftm {L. Maharana, A. Nath and A.R. Panda, Mod. Phys. Lett. 7, 
2275 (1992)}

\def \van {R. Van Royen and V.F. Weisskopf, Nuov. Cim. 51A, 617 (1965)}

\def \rchpi {S.R. Amendolia {\it et al}, Nucl. Phys. B277, 168 (1986)}

\def \chrl{ Y. Nambu, {\PRL{4}{380}{1960}};
A. Amer, A. Le Yaouanc, L. Oliver, O. Pene and
J.C. Raynal,{\PRL{50}{87}{1983a}};{\em ibid}
{\PRD{28}{1530}{1983}};
M.G. Mitchard, A.C. Davis and A.J.
MAacfarlane, {\NPB{325}{470}{1989}};
B. Haeri and M.B. Haeri,{\PRD{43}{3732}{1991}};
V. Bernard,{\PRD{34}{1604}{1986}};
 S. Schramm and
W. Greiner, Int. J. Mod. Phys. \zbf E1, 73 (1992), 
J.R. Finger and J.E. Mandula, Nucl. Phys. \zbf B199, 168 (1982),
S.L. Adler and A.C. Davis, Nucl. Phys.\zbf  B244, 469 (1984),
S.P. Klevensky, Rev. Mod. Phys.\zbf  64, 649 (1992).}

\def \spmijp { S.P. Misra, Ind. J. Phys. 61B, 287 (1987)}

\def \feynman {R.P. Feynman and A.R. Hibbs, {\it Quantum mechanics and
path integrals}, McGraw Hill, New York (1965)}

\def \glstn{ J. Goldstone, Nuov. Cim. \zbf 19, 154 (1961);
J. Goldstone, A. Salam and S. Weinberg, Phys. Rev. \zbf  127,
965 (1962)}

\def \anderson {P.W. Anderson, Phys. Rev. \zbf {110}, 827 (1958)}

\def \nambu{ Y. Nambu, Phys. Rev. Lett. \zbf 4, 380 (1960)}

\def\donogh {J.F. Donoghue, E. Golowich and B.R. Holstein, {\it Dynamics
of the Standard Model}, Cambridge University Press (1992)}

\def\satz {T. Matsui and H. Satz, Phys. Lett. B178, 416 (1986)}

\def\cps {C. P. Singh, Phys. Rep. 236, 149 (1993)}

\def\prliop {A. Mishra, H. Mishra, S.P. Misra, P.K. Panda and Varun
Sheel, Int. J. of Mod. Phys. E 5, 93 (1996)}

\def\hmcor {V. Sheel, H. Mishra and J.C. Parikh, Phys. Lett. B382, 173
(1996); {\it biid}, to appear in Int. J. of Mod. Phys. E}
\def\cort { V. Sheel, H. Mishra and J.C. Parikh, Phys. ReV D59,034501 (1999);
{\it ibid}Prog. Theor. Phys. Suppl.,129,137, (1997).}

\def\surcor {E.V. Shuryak, Rev. Mod. Phys. 65, 1 (1993)} 

\def\stevenson {A.C. Mattingly and P.M. Stevenson, Phys. Rev. Lett. 69,
1320 (1992); Phys. Rev. D 49, 437 (1994)}

\def\mac {M. G. Mitchard, A. C. Davis and A. J. Macfarlane,
 Nucl. Phys. B 325, 470 (1989)} 
\def\tfd
 {H.~Umezawa, H.~Matsumoto and M.~Tachiki {\it Thermofield dynamics
and condensed states} (North Holland, Amsterdam, 1982) ;
P.A.~Henning, Phys.~Rep.253, 235 (1995).}
\def\amph4{Amruta Mishra and Hiranmaya Mishra,
{\JPG{23}{143}{1997}}.}

\def \neglecor{M.-C. Chu, J. M. Grandy, S. Huang and 
J. W. Negele, Phys. Rev. D48, 3340 (1993);
ibid, Phys. Rev. D49, 6039 (1994)}

\def\revdata {Particle Data Group, Phys. Rev. D 50, 1173 (1994)}

\def\sinp {S.P. Misra, Indian J. Phys., {\bf 70A}, 355 (1996)}
\def\npa{H. Mishra and J.C. Parikh, {\NPA{679}{597}{2001}.}}
\def\krisch {M. Alford and K. Rajagopal, JHEP 0206,031,(2002)}
\def\reddy {A.W. Steiner, S. Reddy and M. Prakash,
{\PRD{66}{094007}{2002}.}}
\def\hmam {Amruta Mishra and Hiranmaya Mishra,
{\PRD{69}{014014}{2004}.}}
\def\hmampp {Amruta Mishra and Hiranmaya Mishra,
in preparation.}
\def\bryman {D.A. Bryman, P. Deppomier and C. Le Roy, Phys. Rep. 88,
151 (1982)}
\def\thooft {G. 't Hooft, Phys. Rev. D 14, 3432 (1976); D 18, 2199 (1978);
S. Klimt, M. Lutz, U. Vogl and W. Weise, Nucl. Phys. A 516, 429 (1990)}
\def\alkz { R. Alkofer, P. A. Amundsen and K. Langfeld, Z. Phys. C 42,
199(1989), A.C. Davis and A.M. Matheson, Nucl. Phys. B246, 203 (1984).}
\def\sarah {T.M. Schwartz, S.P. Klevansky, G. Papp,
{\PRC{60}{055205}{1999}}.}
\def\wil{M. Alford, K.Rajagopal, F. Wilczek, {\PLB{422}{247}{1998}};
{\it{ibid}}{\NPB{537}{443}{1999}}.}
\def\sursc{R.Rapp, T.Schaefer, E. Shuryak and M. Velkovsky,
{\PRL{81}{53}{1998}};{\it ibid}{\AP{280}{35}{2000}}.}
\def\pisarski{
D. Bailin and A. Love, {\PR{107}{325}{1984}},
D. Son, {\PRD{59}{094019}{1999}}; 
T. Schaefer and F. Wilczek, {\PRD{60}{114033}{1999}};
D. Rischke and R. Pisarski, {\PRD{61}{051501}{2000}}, 
D. K. Hong, V. A. Miransky, 
I. A. Shovkovy, L.C. Wiejewardhana, {\PRD{61}{056001}{2000}}.}
\def\leblac {M. Le Bellac, {\it Thermal Field Theory}(Cambridge, Cambridge University
Press, 1996).}
\def\bcs{A.L. Fetter and J.D. Walecka, {\it Quantum Theory of Many
particle Systems} (McGraw-Hill, New York, 1971).}
\def\alexander{Aleksander Kocic, Phys. Rev. D33, 1785,(1986).}
\def\bubmix{F. Neumann, M. Buballa and M. Oertel,
{\NPA{714}{481}{2003}.}}
\def\kunihiro{M. Kitazawa, T. Koide, T. Kunihiro, Y. Nemeto,
{\PTP{108}{929}{2002}.}}
\def\igor{Igor Shovkovy, Mei Huang, {\PLB{564}{205}{2003}}.}
\def\prasanth{P. Jaikumar and M. Prakash,{\PLB{516}{345}{2001}}.}
\def\igorr{Mei Huang, Igor Shovkovy, {\NPA{729}{835}{2003}}.}
\def\abrikosov{A.A. Abrikosov, L.P. Gorkov, Zh. Eskp. Teor.39, 1781,
1960}
\def\krischprl{M.G. Alford, J. Berges and K. Rajagopal,
 {\PRL{84}{598}{2000}.}}
\def\hatmampp{A. Mishra and H.Mishra, in preparation}
\def\blaschke{D. Blaschke, M.K. Volkov and V.L. Yudichev,
{\EPJA{17}{103}{2003}}.}
\def\mei{M. Huang, P. Zhuang, W. Chao,
{\PRD{65}{076012}{2002}}}
\def\bubnp{M. Buballa, M. Oertel,
{\NPA{703}{770}{2002}}.}
\def\sarma{G. Sarma, J. Phys. Chem. Solids 24,1029 (1963).}
\def\ebert {D. Ebert, H. Reinhardt and M.K. Volkov,
Prog. Part. Nucl. Phys.{\bf 33},1, 1994.}
\def\rehberg{ P. Rehberg, S.P. Klevansky and J. Huefner,
{\PRC{53}{410}{1996}.}}
\def\lutz{M. Lutz, S. Klimt, W. Weise,{\NPA{542}{521}{1992}.}}
\def\rapid{B. Deb, A.Mishra,H. Mishra and P. Panigrahi,
Phys. Rev. A {\bf 70},011604(R), 2004.}
\def\kriscfl{M. Alford, C. Kouvaris, K. Rajagopal, Phys. Rev. Lett.
{\bf 92} 222001 (2004), arXiv:hep-ph/0406137.}
\def\shovris{S.B. Ruester, I.A. Shovkovy and D.H. Rischke,
arXiv:hep-ph/0405170.}
\def\krisaug{K. Fukushima, C. Kouvaris and K. Rajagopal, arxiv:hep-ph/0408322}.
\def\wilczek{W.V. Liu and F. Wilczek,{\PRL{90}{047002}{2003}},E. Gubankova,
W.V. Liu and F. Wilczek, {\PRL{91}{032001}{2003}.}}
\def\review{For reviews see K. Rajagopal and F. Wilczek,
arXiv:hep-ph/0011333; D.K. Hong, Acta Phys. Polon. B32,1253 (2001);
M.G. Alford, Ann. Rev. Nucl. Part. Sci 51, 131 (2001); G. Nardulli,
Riv. Nuovo Cim. 25N3, 1 (2002); S. Reddy, Acta Phys Polon.B33, 4101(2002);
T. Schaefer arXiv:hep-ph/0304281; D.H. Rischke, Prog. Part. Nucl. Phys. 52,
197 (2004); H.C. Ren, arXiv:hep-ph/0404074; M. Huang, arXiv: hep-ph/0409167;
I. Shovkovy, arXiv:nucl-th/0410191.}
\def\kunihiroo{ M. Kitazawa, T. Koide, T. Kunihiro and Y. Nemoto,
{\PRD{65}{091504}{2002}}, D.N. Voskresensky, arXiv:nucl-th/0306077.}
\def\rupak{S.Reddy and G. Rupak, arXiv:nucl-th/0405054}

\end{document}